\documentclass[conference]{IEEEtran}

\usepackage[utf8]{inputenc}

\IEEEoverridecommandlockouts
\usepackage{cite}
\usepackage{amsmath,amssymb,amsfonts}

\usepackage{algorithmic,xpatch}

\usepackage{algorithm}
\usepackage{etoolbox}

\usepackage{url}
\usepackage{multirow}

\usepackage{tabularx}
\usepackage{hyphenat}

\usepackage{array}

\usepackage{graphicx}
\usepackage{textcomp}
\usepackage{xcolor}
\def\BibTeX{{\rm B\kern-.05em{\sc i\kern-.025em b}\kern-.08em
    T\kern-.1667em\lower.7ex\hbox{E}\kern-.125emX}}

\makeatletter

\xpatchcmd{\algorithmic}
  {\newcommand{\STATE}{\ALC@it}}
  {\newcommand{\STATE}{\@ifstar\STATEstar\STATEnostar}}
  {}{}
\newcommand{\STATEstar}{\item[]}
\newcommand{\STATEnostar}{\ALC@it}
\makeatother

\begin{document}

\title{
A Bag-of-Tasks Scheduler  Tolerant to   Temporal Failures in  Clouds
}

\author{\IEEEauthorblockN{ Luan Teylo and  L\'{u}cia Maria de A. Drummond}
\IEEEauthorblockA{\textit{  Fluminense Federal University} \
\textit{Institute of Computing}\\
Niteroi, Brazil \\
(luanteylo, lucia)@ic.uff.br}
\and
\IEEEauthorblockN{Luciana Arantes and Pierre Sens}
\IEEEauthorblockA{\textit{Sorbonne Universit\'{e}, CNRS, INRIA} \\
LIP6, Paris, France\\
(luciana.arantes, pierre.sens)@lip6.fr}
}

\maketitle

\begin{abstract}

Cloud platforms have emerged as a prominent environment to execute high performance computing (HPC) applications providing on-demand resources as well as scalability. They usually offer different classes of Virtual Machines (VMs) which ensure different guarantees in terms of availability and volatility, provisioning the same resource through multiple pricing models. For instance, in Amazon EC2 cloud, the user pays per hour for {\it on-demand} VMs while  {\it spot} VMs are unused  instances available for lower price. Despite the monetary advantages,  a spot VM  can be terminated, stopped, or hibernated by EC2 at any moment.   

Using both hibernation-prone spot VMs (for cost sake) and on-demand VMs, we propose in this paper a static scheduling for HPC applications which are composed by independent tasks (bag-of-task) with deadline constraints. However, if a spot VM hibernates and it does  not resume within a time which guarantees the application's  deadline, a temporal failure takes place. Our scheduling, thus, aims at  minimizing monetary costs of bag-of-tasks applications  in EC2 cloud, respecting its deadline and avoiding temporal failures.  To this end, our algorithm statically creates two scheduling maps: (i) the first one contains, for each task, its starting time and on which VM (i.e., an available spot or on-demand VM with the current lowest price) the task should execute; (ii) the second one contains, for each task allocated on a VM spot in the first map,  its starting time and on which on-demand VM it should be executed to meet the application deadline in order to avoid temporal failures. The latter will be used whenever the hibernation  period of a spot VM exceeds a  time limit. 
 
Performance results from simulation with task execution traces, configuration of Amazon EC2 VM classes, and  VMs market history confirm the effectiveness of our scheduling and that it tolerates temporal failures.

\end{abstract}


\section{Introduction}

High Performance Computing (HPC) applications are typically executed in dedicated data centers. However, in the past few years, cloud computing has emerged as an attractive option to run these applications due to several advantages that it brings when compared with a dedicated infrastructure.
Clouds provide a significant reduction in operational costs, besides offering a rapid elastic provisioning of computing resources like virtual machines and storage. However, in cloud environments,  besides the usual  goal of minimizing the execution time of the HPC application,  it is also important to minimize the monetary cost of using cloud resources, i.e., there exists a trade-off between performance and monetary cost.

In this paper, we are interested in HPC bag-of-task (BoT) applications with time constraints (deadlines) within which they must finish. BoT applications are composed of independent tasks which can be executed in any order and in parallel.   Although simple, the BoT approach is used by several HPC applications such as parameter sweep applications, chromosome mapping, Monte Carlo simulation, computer imaging applications~\cite{SmallenCB01}, \cite{CasanovaOBW00},\cite{MascagniL03},\cite{WhiteTW90}. Furthermore, they may require deadline-bounds where the correctness on the computation also depends on the time the computation of all tasks ends.

Infrastructure-as-a-Service (IaaS) existing cloud platforms (e.g., Amazon EC2, Microsoft Azure, Google Cloud, etc.) enable users to dynamically acquire resources, usually as virtual machines (VMs), according to their application requirements (CPU, memory, I/O, etc,) in a pay-as-you-use price model.  They usually offer different classes of VMs which ensure different guarantees in terms of availability and volatility, provisioning the same resource through multiple pricing models. For instance, in Amazon EC2, there are basically three classes\footnote{\url{https://aws.amazon.com/ec2/instance-types/}}: (i) {\it reserved} VM instances, where the user pays an upfront price, guaranteeing long-term availability; (ii) {\it on-demand} VM instances which are allocated for specific time periods and incur a fixed cost per unit time of use, ensuring availability of the instance during this period; (iii) {\it spot} VM instances which are an unused instances available for lower price than on-demand price.

The availability of spot VMs instances fluctuates based on the spot market's current demand. The allocation of a spot instance involves defining the VM type and a maximum price for how much the user is willing to pay. However, if there are not enough instances to meet clients demands, the VM in question can be interrupted by the cloud provider  (temporarily or definitively). Despite the risk of unavailability, the main advantage of spot VMs is that their cost is much lower than on-demand VMs since the user requests unused instances at steep discounts, reducing the costs significantly.

With Amazon’s more recent announcement, an interrupted spot can either terminate, stop, or hibernate. 
Hence, when requesting a  spot instance the user specifies the required type as well as the action that Amazon EC2 should take in case the VM instance is interrupted.  Whenever a spot instance is hibernated by EC2, its memory and context are saved to the root of EC2 Block Storage (EBS) volume and, during the VM's pause, the user is only charged for EBS storage. EC2 resumes the hibernated instance, reloading the saved memory and context, only when there are enough availability for that type of instance with a spot price which is lower than the user's maximum price. Contrarily  to stopped or terminated instances whose user is warned two minutes before the interruption of them, hibernated instances are paused immediately after noticing the user.

 Our proposal in this work is to provide a static cloud scheduler for Bag-of-Tasks applications using, for cost sake, hibernate-prone spot instances as much as possible, respecting the application deadline constraints while also minimizing the  monetary costs of bag-of-tasks applications. However, if a spot instance hibernates, it might happen that it will not resume within a time which guarantees the deadline constraints of the application. In this case, a {\it temporal failure} would take place, i.e., correct computation is performed but too late to be useful (inability to meet deadlines). Thus, in order to avoid temporal failure in case of spot instance hibernation, our scheduler statically computes the time interval that an hibernated instance can stay in this state without violating the application's deadline. If the instance does not resume till the end of this interval, our scheduler will move the execution of the current tasks of the spot instance as well as those not executed yet to on-demand instances, in order to guarantee the application's deadline. Note that even after migrating the remaining task execution to on-demand VMs, the scheduler continues to look forward to minimizing  monetary costs. 
 
 The rest of the paper is organized as follows. Section~\ref{sec:related_work} discusses some related work. Section~\ref{sec:scheduler} describes our proposed static scheduling, including its algorithms. Evaluation results from simulations conducted with real traces are presented in section \ref{sec:results}. Finally, Section~\ref{sec:conclusion} concludes the paper and presents some future directions.

\section{Related Work}
\label{sec:related_work}

Bag-of-tasks on  clouds are widely used not only for scientific applications but also for many commercial applications. In \cite{GSW2015}, Facebook reports that the jobs running on their own internal data centers are mostly independent tasks. Many works propose then scheduling the execution of independent tasks both on homogeneous and heterogeneous cloud environments \cite{ThaiVB18}. In the former, the performance and pricing of all available VMs are the same. In this case, authors usually consider either reserved VMs \cite{YaoZlHLL14} or on-demand VMs \cite{ThaiVB14}. For instance, Thai et al. \cite{ThaiVB14} study scheduling of applications on on-demand VMs distributed across different datacenters, focusing on the trade-offs between performance and cost while Yao et al. \cite{YaoZlHLL14} provides a solution that satisfies job deadlines while minimizing monetary cost. The proposed heuristics use both on-demand and reserved VMs. Works on heterogeneous cloud consider different types of VMs. For instance, in \cite{ThaiVB15} the authors present a heuristic algorithm for executing a bag-of-tasks applications taking into account either budget or deadline constraints. In \cite{ThaiVB18}, Thai et al. present an extensive survey and taxonomy of existing research in scheduling of bag-of-task applications on clouds.


Several works propose to tolerate failures (crash) of VMs in Clouds using checkpointing, passive or active replication, or task resubmission approaches. Some of them, similarly to our approach, must avoid job and/or application's deadline violation.

Checkpointing \cite{GoiriJGT10, AupyBMRR13} approaches periodically save the execution state of a VM as an image file. These mechanisms are costly since data centers have limited network resources and may readily become overloaded when a huge number of checkpoint image files need to be stored.  Moreover, they are not suitable for real-time context or jobs with deadline constraints since periodically checkpoints and resume of failed tasks are time-consuming.

Replication and resubmission of tasks are the other mechanisms widely used to tolerate failures \cite{PlankensteinerPFKK08, ZhengVT09, WangBZYX15}.  Using replication, several copies of the same task are executed to support fault tolerance. Most of studies use a single primary-backup scheme considering one primary and one or several backup (copy) tasks scheduled on different computing instances \cite{ZhengVT09, WangBZYX15}. The copies of a task are executed only when the primary task fails. In order to reduce the response time in case of failure, overlapping techniques are proposed \cite{ZhengVT09} in a Grid context where a backup is scheduled for each primary on a different host.
A backup is executed when its primary cannot complete execution due to a failure but it does not require fault diagnosis. Zheng et al. \cite{ZhengVT09} propose an algorithm to find an optimal backup schedule for each independent tasks.  Wang et al. \cite{WangBZYX15} extend Zheng's results in a cloud context and using an elastic resource provisioning.

Despite the use of overlapping techniques, primary-backup schemes require that the tasks deadlines have enough time for executing backups in case of failure. Then, several works study active replication \cite{ Al-OmariSM04,CirneBSGV07, BenoitHR08} allowing backups to execute concurrently with its primaries. Al-Omari et al. \cite{Al-OmariSM04} improve the primary-backup scheme by proposing the primary-backup-overloading technique, in which the primary of a task can be scheduled onto the same or overlapping time interval with the backup of another task on a processor. In \cite{CirneBSGV07}, authors present a comprehensive study of replication schedulers where all replicas of a task start executing concurrently and the next task is started as soon as one of the previous task replicas finish. Benoit et al. \cite{BenoitHR08} adopt a more conservative approach where the next task can only start when all the replicas of the previous task finished.

Contrarily to our approach, the above solutions need to schedule both the primary and backup tasks and the latter take the execution control if the former fail. Neither of them use backup tasks to avoid temporal failure. In addition, in our case, they are only executed in case of VM's hibernation and risk of temporal failures.

Some works take into account Amazon spot VMs instance features. In \cite{LuLWKPHLL13}, Lu et al. use hybrid instances, including both on-demand instances for high priority tasks and backup, and spot instances for normal computational tasks.  Authors of \cite{MenacheSJ14} propose to switch to on-demand resources when there is no spot instance available to ensure the desired performance. Using both on-demand and spot VM instances, SpotCheck \cite{SharmaLGIS15} provides the illusion of an IaaS platform that offers always-available on-demand VMs for a cost near that of spot VMs.  Also claiming performance
of on-demand VMs, but at a cost near that of
the spot market,  the authors in \cite{SubramanyaGSIS15} present the SpotOn batch service computing, that uses fault-tolerance mechanism to mitigate the impact of spot revocations.
To our knowledge no work studies the impact of the new hibernation feature of spot instances on scheduling algorithms.

\section{A Static Scheduler of Bag-of-tasks Applications  in  Clouds}
\label{sec:scheduler}

Aiming at reducing monetary costs, our proposed scheduling uses hibernate-prone spot instances. However, due to the possibility of hibernation  and also the need to meet the application's deadline,  the scheduler might migrate tasks that run on spot instances to on-demand ones, whenever the duration of an instance  hibernation would induce a  temporal failure.  We denote  {\it primary} tasks  those which are allocated on VMs (spot or on-demand) that guarantee application's deadline with minimum monetary cost and we denote {\it backup} tasks those which are allocated on on-demand VMs and were originally primary tasks allocated  on spot VMs. Backup tasks are only executed in case the hibernation state remains for such a long period of time that it is impossible to meet the deadline of the application, avoiding, thus, temporal failures.  Therefore, a task might have two versions (primary and backup) which are statically scheduled on two different cores with time exclusion. 

The scheduling outputs two allocation mappings: one with primary tasks and the other one with backup tasks.

Concerning the primary mapping,  the proposed  strategy aims at minimizing the monetary costs, by adopting hibernate-prone spot instances with the highest processing power.
Regarding the backup mapping, our strategy aims at minimizing monetary costs, by  using the minimum   number of  the cheapest on-demand VMs,  without violating the application's deadline.


We assume that each task of the BoT application is executed in one core, requiring some main memory and that a set of different types of VMs are  usually offered by cloud providers with a varying  number of virtual cores (VCPUs) and memory sizes. Therefore, a VM running on a multi-core machine can execute  more than one task simultaneously (one VCPU per task) provided there is enough main memory to allocate them. We also consider that VMs are offered in  two different markets, spot and on-demand, where, contrarily to the former, the latter can not hibernate. Note that our solution only allocates spot VMs of those types that support hibernation.

\begin{figure}
    \centering
    \includegraphics[scale=0.35]{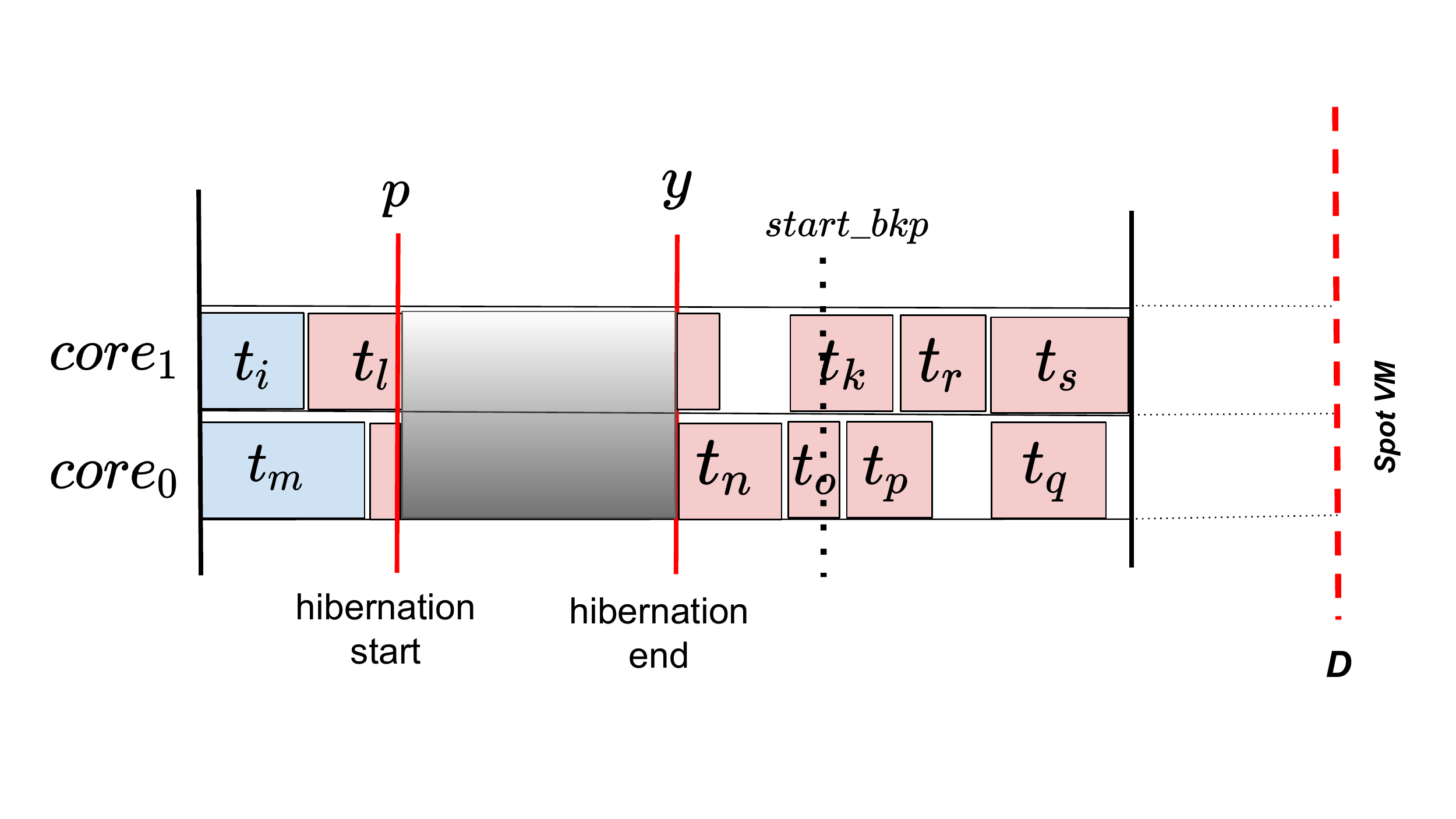}
    \caption{Hibernation without Backup Execution.}
    \label{fig:hib1}
\end{figure}

\begin{figure}
    \centering
    \includegraphics[scale=0.4]{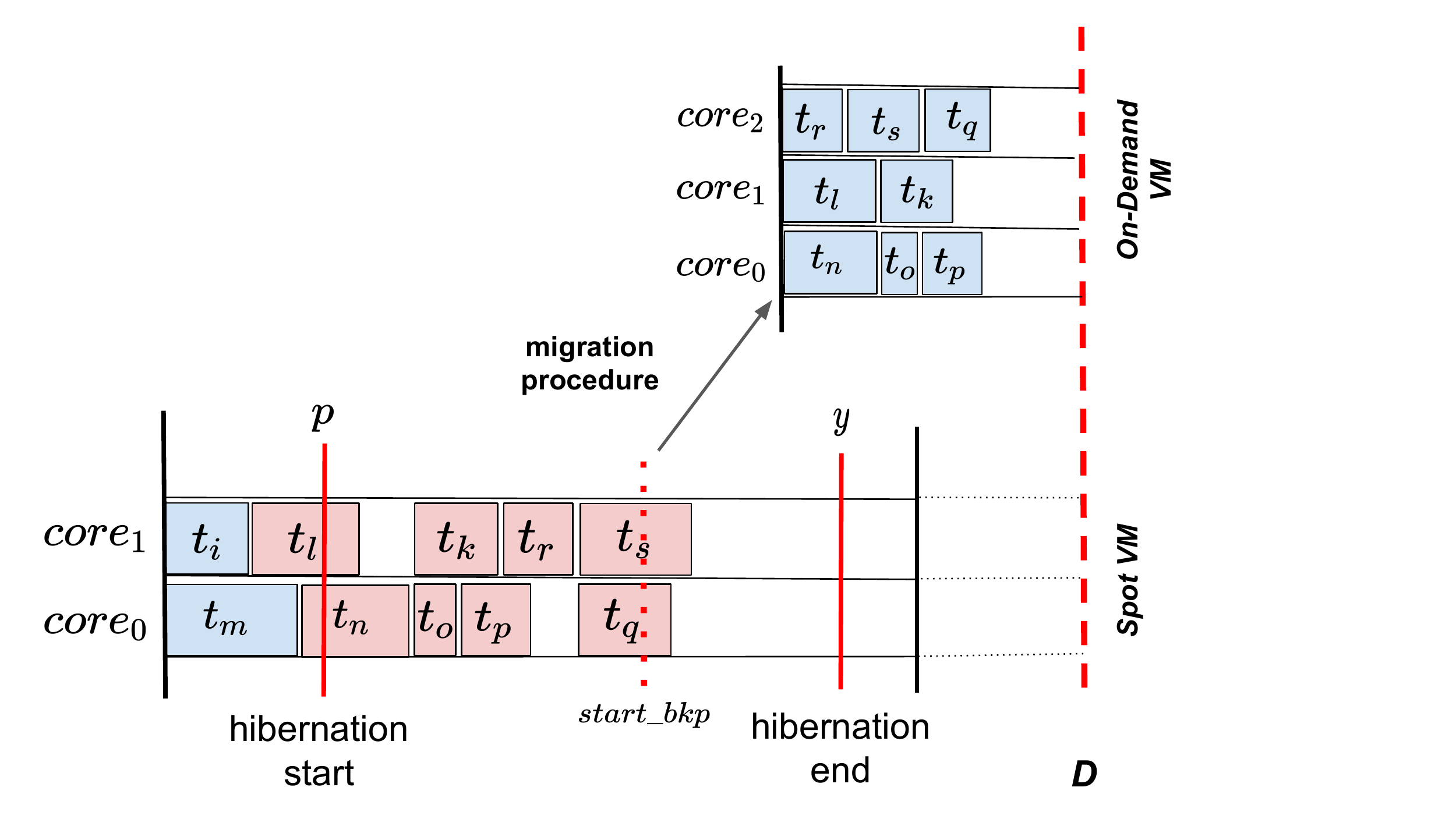}
    \caption{Hibernation with Backup Execution. }
    \label{fig:hib2}
\end{figure}

Figure \ref{fig:hib1} shows an example  where the hibernation does not require backup tasks execution. In this example,  a spot instance starts hibernating in time $p$ and finishes in $y$, before the time limit, $start\_bkp$, when   the backups should be triggered. Then, the deadline $D$ can be met without executing the backups. On the other hand,  Figure \ref{fig:hib2} presents a case where it is necessary to execute the backup tasks in an on-demand virtual machine to meet the deadline, since the hibernation exceeded the time limit, $start\_bkp$.

Let $M$ be the set of virtual machines, $B$ the set of tasks that compose a bag-of-task application, and $T = \{1, \ldots, D\}$ the set of feasible periods, where $D$ is the deadline defined by the user. 
For each VM,  $M$ keeps its storage capacity and the number of cores with the corresponding computation power. Set $B$ keeps, for each task, information about (i) its execution time on a machine with known computational power (base time duration) and (ii) the amount of  main memory that the task needs. Let $Queue^{vm_j} \subset B$ be the set with all tasks scheduled on $vm_j$.

It is worth mentioning that the execution time of a task is re-calculated as the product between the original  execution  time and the VM slowdown where it will be executed. A  VM slowdown is defined  as  $\frac{P_B}{P_{vm_{j}}}$, where $P_B$ is the  processing capacity of the machine  used to calculate the basis time, and   $P_{vm_{j}}$  is the  processing capacity of the VM. Thus,  the  slowdown represents the processing capacity of a VM when compared with the  machine used to compute the basis  time duration.

When a  VM is allocated for a user, he/she pays for a full-time interval called $slot$. That time is usually one hour. Thus,  if a VM is used for 61 minutes,  for example, the user will be charged for two $slots$ (120 minutes).  Note that one slot can correspond to several periods. For example, if each period corresponds to one minute, a slot of  one hour  would correspond to 60 periods.   It is, thus, in the user's best interest to maximize the use of a slot already allocated.

Let  $start\_slot^{vm_j}$  and  $end\_slot^{vm_j}$ be the time when the first slot was allocated to $vm_j$ and the end time of the last allocated slot for this same VM respectively, such that \allowbreak $start\_slot^{vm_j} \allowbreak < end\_slot^{vm_j}$.
Whenever the execution time of a task allocated to $vm_j$ exceeds the $end\_slot^{vm_j}$, the user has to pay for another full interval.
Thus, if part of that interval is not used by any task, we  have a waste of time. To compute that waste, we define  $waste^{vm_j}$ in Equation \ref{eq:wast},  that is the time interval inside the last contracted $slot$ at which $vm_j$ remains idle after executing all tasks allocated to it.

\begin{equation}
\label{eq:wast}
    \begin{split}
        &waste^{vm_j} = end\_slot^{vm_j}- end_{t_{max}}\\
    \end{split}
\end{equation}

Such that $end_{t_{max}} = \max\limits_{\forall t_l \in Queue^{vm_j}}(end_{t_l}) $ and $end_{t_{l}}$ is the end time of task  $t_{l}$.

\subsection{Primary Task Scheduling}

Algorithm \ref{alg:static} shows the primary scheduling heuristic which is a greedy algorithm that allocates the set of tasks $t_i \in B$ to a set of VMs (spot and on-demand VMs). Tables \ref{tab:var1} and  \ref{tab:proc1} present the used variables and functions respectively.
The algorithm receives  $B$, $M$, $D$, and $VM\_time_{limit}$ as input parameters. The $VM\_time_{limit}$ defines the maximum occupation period of a VM. For example, if $D = 100 (h)$ and $VM\_time_{limit} = 0.5$,  the scheduling of the tasks should be done so as not to exceed the period $ D * VM\_time_{limit} = 50 (h)$.
Since the objective is to respect the application  deadline (even in the presence of hibernation) while minimizing monetary costs, all the choices made by the heuristic are guided by the VMs' prices, and by the deadline $D$ and  $VM\_time_{limit}$, defined by the user.

Initially, tasks are ordered in descending order by the memory size they require (line \ref{algP:sort}). Then, for each task, the algorithm applies a best fit heuristic that tries to include it in an already allocated slot of a virtual machine that presents the highest $waste$ of time (lines \ref{algP:vmfor} to \ref{algP:endvmfor}), since it has enough memory and ensures that the task insertion will respect $D * VM\_time_{limit}$.
If such a VM does not exist, the heuristic tries to allocate new slots in an already  allocated  VM with enough memory to execute the task,  but  now with the smallest $waste$  (lines \ref{algP:nslot} to \ref{algP:endnslot}). Similarly to the previous case, the slot allocation must not violate $D * VM\_time_{limit}$ (line \ref{ifslotdeadline}).

Allocating slots in an already allocated VM reduces boot time overhead in comparison of allocating a new VM.  However, if such an allocation is not possible, the algorithm must allocate  a new VM. In this case, the heuristic  defines the best type of VM in terms of execution time (line \ref{algP:bestVM}) and, then, it chooses the  market where this VM shall be acquired: on-demand or spot, considering the offered prices (lines \ref{algP:market} to \ref{algP:endmarket}).
Finally,  it updates the primary scheduling map (line \ref{AlgP:update}).

{
\begin{algorithm}[]
		\caption{\textit{Primary Task Scheduling}}
		\label{alg:static}
		\small
        \begin{algorithmic}[1]
		\renewcommand{\algorithmicrequire}{\textbf{Input:}}
		\renewcommand{\algorithmicensure}{\textbf{Output:}}
        \REQUIRE $B$, $M$, $D$, $VM\_time_{limit}$
		\STATE $sort(B)$; \label{algP:sort}
	    \STATE $PQ = \emptyset$;
		\STATE $A\_VM = \emptyset$; { /* set of allocated VMs */}
			\FOR{$ \textbf{ all } t_i \in B$} \label{algP:mainfor}
			    \STATE $sort\_by\_max\_waste(A\_VM)$; {/* using Equation \ref{eq:wast} */}  
			    \STATE $inserted = False$;
			    
			    \STATE* \textit{/*Check if $vm_k$  has sufficient time and memory in an  already allocated  slot to execute $t_i$ without violating the limit $D*VM\_time_{limit}$*/}
			    \FOR{$ \textbf{ all }  vm_k \in A\_VM$}\label{algP:vmfor}
			    \IF{$check\_insertion(t_i, vm_k, D*VM\_time_{limit})$}
			        \STATE $insert(t_i, vm_k)$; 
			        \STATE $inserted = True$;
			        \STATE $break$;
			    \ENDIF
			   \ENDFOR\label{algP:endvmfor}
			
			\STATE* \textit{/*Check if it will be necessary to allocate a new slot on an already allocated VM or if it will be necessary to allocate a new VM*/}
			\IF{$\textbf{ not } inserted$}
			    \STATE $vm_{aux} = NONE$;
			    \STATE $sort\_by\_min\_wast(A\_VM)$; \label{algP:nslot}
			    \FOR{ $ \textbf{ all }  vm_k \in A\_VM \textbf{ with enough memory}$}
			        \STATE* \textit{/*get the number of slots necessary to execute $t_i$ on $vm_k$*/}
			        \STATE $n =  number\_of\_slots(t_i, vm_k)$; 
			        \IF{$end\_slot^{vm_k} + (n *  slot) < D*VM\_time_{limit}$}\label{ifslotdeadline}
			            \STATE $vm_{aux} = vm_k$;
			            \STATE $break$;
			        \ENDIF
			    \ENDFOR\label{algP:endnslot}
			   
			    \IF{$ vm_{aux} \textbf{ is equal to } NONE $}
			        \STATE $vm_{aux} = best\_VM(t_i, M)$;\label{algP:bestVM}
			        \IF{$spot\_price^{vm_{aux}} < odm\_price^{vm_{aux}}$} \label{algP:market}
			             \STATE $vm_{aux}^{market}= spot$;
			        \ELSE
			             \STATE $vm_{aux}^{market} = \textit{on-demand}$;
			        \ENDIF \label{algP:endmarket}
			        
			    \ENDIF
			   
			    \STATE $n =  number\_of\_slots(t_i, vm_{aux})$;
			    \STATE*  \textit{/*Allocate  the number of slots required to execute $t_i$ in $vm_{aux}$*/}
			    \STATE $allocate\_slots(vm_{aux}, n)$;
			    \STATE $insert(t_i, vm_{aux})$
			    \STATE $update(A\_VM)$;{ \textit{/*Update the set of allocated VMs*/}} \label{AlgP:update}
		   \ENDIF
	\ENDFOR
	\STATE $ PQ = create\_primary\_map(A\_VM)$;
	\end{algorithmic}
\end{algorithm}
 }

\begin{table}[htbp]
\caption{Variables of Primary Scheduling Heuristic \ref{alg:static}}
\begin{center}
\centering\begin{tabular*}{\linewidth}{|c|l|l}
\hline
\textbf{Name} & \multicolumn{2}{|l|}{\textbf{\textit{Description}}} \\
\hline
\begin{tabular}[c]{@{}l@{}}$B$\end{tabular} & \multicolumn{2}{p{5.8cm}|}{Set of tasks} \\
\hline
\begin{tabular}[c]{@{}l@{}}$M$ \end{tabular} & \multicolumn{2}{p{5.8cm}|}{Set of  VMs} \\
\hline
\begin{tabular}[c]{@{}l@{}}$D$\end{tabular} & \multicolumn{2}{p{5.8cm}|}{Deadline defined by the user, to be respected even in presence of VM hibernation}\\
\hline
\begin{tabular}[c]{@{}l@{}}$VM\_time_{limit}$\end{tabular} & \multicolumn{2}{p{5.8cm}|}{Parameter that determines the maximum occupation period of a VM }\\
\hline
\begin{tabular}[c]{@{}l@{}}$A\_VM$\end{tabular}& \multicolumn{2}{p{5.8cm}|}{Set of VMs selected to execute primary tasks} \\
\hline
\begin{tabular}[c]{@{}l@{}}$t_i$ \end{tabular} & \multicolumn{2}{p{5.8cm}|}{Task $i$} \\
\hline
\begin{tabular}[c]{@{}l@{}}$inserted$\end{tabular} & \multicolumn{2}{p{5.8cm}|}{Boolean variable that indicates whether task $t_i$ was successfully scheduled} \\
\hline
\begin{tabular}[c]{@{}l@{}}$vm_k, vm_{aux}$\end{tabular} & \multicolumn{2}{p{5.8cm}|}{Virtual machines} \\
\hline
\begin{tabular}[c]{@{}l@{}}$slot$\end{tabular} & \multicolumn{2}{p{5.8cm}|}{Minimum contracted time for a VM (for example, in AWS the slot is 1 hour) } \\
\hline
\begin{tabular}[c]{@{}l@{}}$n$\end{tabular} & \multicolumn{2}{p{5.8cm}|}{Number of contracted slots} \\
\hline
\begin{tabular}[c]{@{}l@{}}$end\_slot^{vm_k}$ \end{tabular} & \multicolumn{2}{p{5.8cm}|}{End of last contracted slot of  $vm_k$} \\
\hline
\begin{tabular}[c]{@{}l@{}}$vm_{aux}^{market}$\end{tabular} & \multicolumn{2}{p{5.8cm}|}{Market where $vm_{aux}$ will be contracted: on-demand or spot} \\
\hline
\begin{tabular}[c]{@{}l@{}}$spot\_price^{vm_k}$ \end{tabular}& \multicolumn{2}{p{5.8cm}|}{Price of a slot of $vm_k$ in the spot market}\\
\hline
\begin{tabular}[c]{@{}l@{}}$odm\_price^{vm_{aux}}$\end{tabular} & \multicolumn{2}{p{5.8cm}|}{Price of a slot of $vm_k$ in the on-demand market} \\
\hline
\begin{tabular}[c]{@{}l@{}}$PQ$\end{tabular} & \multicolumn{2}{p{5.8cm}|}{ Scheduling map of primary tasks  containing  VMs  and the corresponding execution queues} \\
\hline
\end{tabular*}
\label{tab:var1}
\end{center}
\end{table}

 \begin{table}[htbp]
\caption{Functions and Procedures of Primary Scheduling Heuristic \ref{alg:static}}
\begin{center}
\begin{tabular*}{\linewidth}{|c|l|l}
\hline
\textbf{Name} & \multicolumn{2}{|l|}{\textbf{\textit{Description}}} \\
\hline
\begin{tabular}[c]{@{}l@{}}$sort(B)$\end{tabular} & \multicolumn{2}{p{4.05cm}|}{Sort the set of tasks $B$ in descending order by  memory size  demand} \\
\hline
\begin{tabular}[c]{@{}l@{}}$sort\_by\_max\_waste(A\_VM)$\end{tabular} & \multicolumn{2}{p{4.05cm}|}{Sort the set of  VMs $A\_VM$ in descending order by the waste size} \\
\hline
\begin{tabular}[c]{@{}l@{}}$sort\_by\_min\_waste(A\_VM)$ \end{tabular} & \multicolumn{2}{p{4.05cm}|}{Sort the set of  VMs $A\_VM$ in ascending order by the waste size} \\
\hline
\begin{tabular}[c]{@{}l@{}}$check\_insertion(t_i, vm_k, D)$\end{tabular} & \multicolumn{2}{p{4.05cm}|}{Check if $vm_j$ has an already contracted slot with enough idle time and memory  to execute task $t_i$,   and  if that insertion respects the deadline $D$} \\
\hline
\begin{tabular}[c]{@{}l@{}}$ insert(t_i, vm_k)$ \end{tabular}& \multicolumn{2}{p{4.05cm}|}{Insert task $t_i$ into the execution queue of $vm_k$} \\
\hline
\begin{tabular}[c]{@{}l@{}}$ number\_of\_slots(t_i, vm_k)$\end{tabular} & \multicolumn{2}{p{4.05cm}|}{Compute the number of  slots necessary to execute task $t_i$ in $vm_k$} \\
\hline
\begin{tabular}[c]{@{}l@{}}$ best\_VM(t_i, M)$ \end{tabular}& \multicolumn{2}{p{4.05cm}|}{Select the VM that executes task $t_i$ with the minimum number of periods of time} \\
\hline
\begin{tabular}[c]{@{}l@{}}$ allocate\_slots(vm_{aux}, n)$\end{tabular} & \multicolumn{2}{p{4.05cm}|}{Allocate (contract) $n$ slots in  $vm_{aux}$} \\
\hline
\begin{tabular}[c]{@{}l@{}}$ update(A\_VM)$ \end{tabular}& \multicolumn{2}{p{4.05cm}|}{Update $A\_VM$ either with  new slots in an already  selected VM or with  the inclusion of  a new VM} \\
\hline
\begin{tabular}[c]{@{}l@{}}$ create\_primary\_map(A\_VM)$ \end{tabular} & \multicolumn{2}{p{4.05cm}|}{Create the  scheduling map of primary tasks} \\
\hline
\end{tabular*}
\label{tab:proc1}
\end{center}
\end{table}

\begin{figure}[!htb]
    \centering
    \includegraphics[scale=0.35]{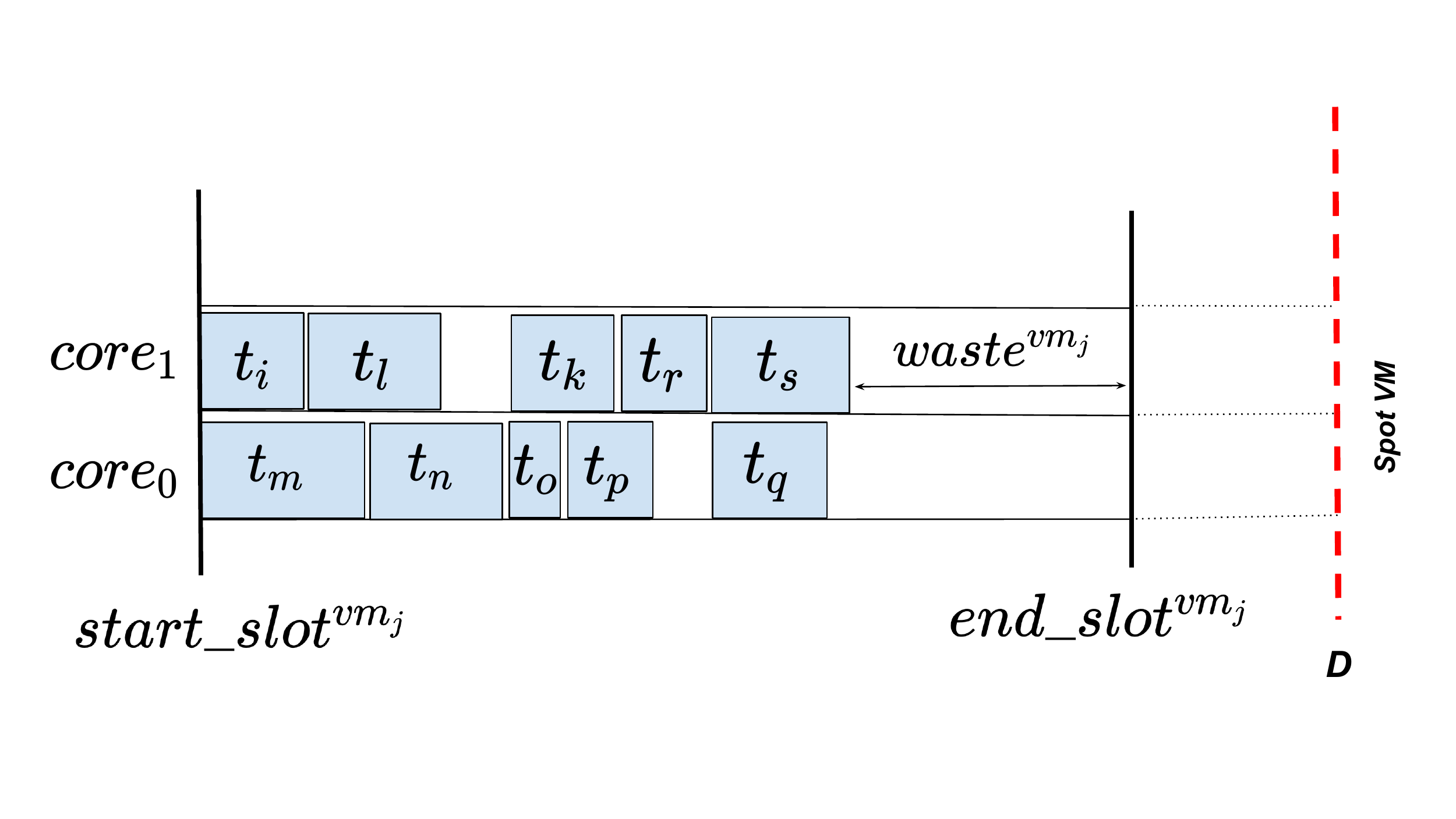}
    \caption{Primary Scheduling  of tasks within a VM slot.}
    \label{fig:pscheduling}
\end{figure}

Figure \ref{fig:pscheduling} shows an example  of  scheduling of nine tasks in a  virtual machine with two cores.  In the example, there exist two gaps (one per core) which  occur due to lack of  memory to allocate a task within the current slot. The  $waste$ of time  and  deadline $D$  are also shown.

\subsection{Backup Task Scheduling}

\begin{figure}[!htb]
    \centering
    \includegraphics[scale=0.35]{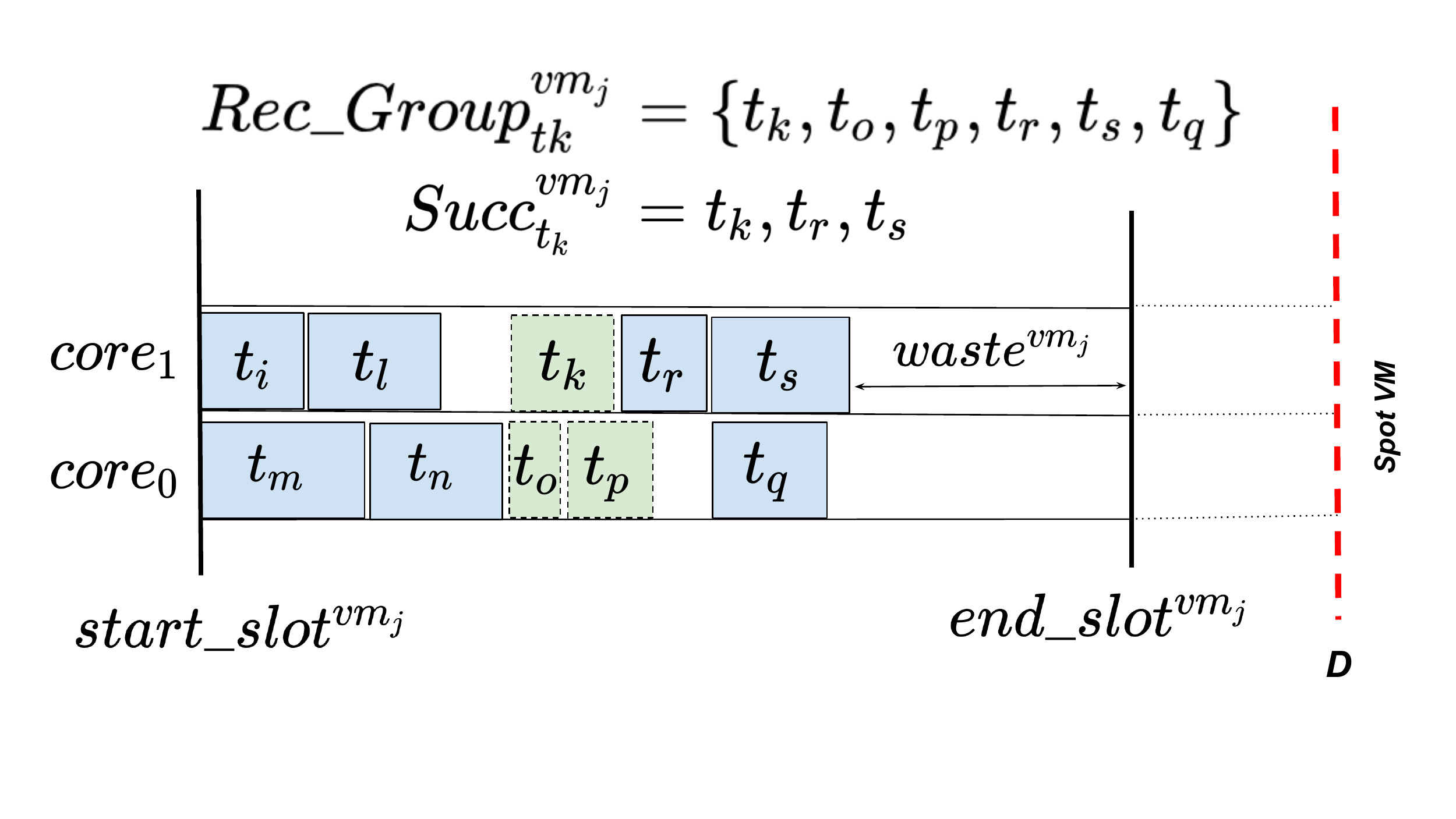}
    \caption{Example of  a task $t_k$,  its successors and which backup tasks might be executed to prevent   $vm_j$  from a temporal failure while executing  task $t_k$  ($Rec\_Group_{t_i}^{vm_j}$).}
    \label{fig:backup}
\end{figure}

Let $Succ_{t_k}^{vm_j} \subset Queue^{vm_j}$ be  a set containing task $t_k$ and all its successors, i.e., all tasks that are allocated to the same core where  $t_k$ is allocated and  that execute after the end of $t_k$.  Let $Parallel_{t_i}^{vm_j} \subset Queue^{vm_j}$ be a set containing all tasks that execute in parallel with $t_i$ in $vm_j$.
In order to avoid temporal failures due to $vm_j$'s hibernation while executing $t_i$ in one of its core, it is necessary to determine which backup tasks must be executed in this case. To this end, we define  $Rec\_Group_{t_i}^{vm_j} \subset Queue^{vm_j}$, as presented  in Equation \ref{eq:error}. The set $Rec\_Group_{t_i}^{vm_j}$ is obtained by the union of all $Succ_{t_k}^{vm_j}$, such that $t_k \in Parallel_{t_i}^{vm_j}$ or $t_k = t_i$. We also define the set $S\_VM_{t_i} \subset M$, that contains all VMs that will be used to execute backup tasks of $Rec\_Group_{t_i}^{vm_j}$, if a migration occurs. Figure \ref{fig:backup} shows an example of   $Succ_{t_k}^{vm_j}$ and  $Rec\_Group_{t_k}^{vm_j}$  sets of task $t_k$ allocated to $vm_j$.

\begin{equation}
\label{eq:error}
    \begin{split}
&Rec\_Group_{t_i}^{vm_j} = \bigcup\limits_{t_k \in (Parallel_{t_i}^{vm_j} \cup \{t_i\})  } Succ_{t_k}^{vm_j}\\
    \end{split}
\end{equation}

  We also define the backup start time,  $start\_bkp^{vm_j}_{t_i}$, as presented in  Equation \ref{eq:bkp}. 
 It defines  how long the hibernation state of  $vm_j$ can be tolerated before any action of migrating tasks of $Rec\_Group_{t_i}^{vm_j}$ to backup ones is  triggered.

 \begin{equation}
 \footnotesize
\label{eq:bkp}
    \begin{split}
    &start\_bkp^{vm_j}_{t_i} = D-runtime(Rec\_Group_{t_i}^{vm_j}, S\_VM_{t_i})-1
    \end{split}
\end{equation}

Such that  $runtime(Rec\_Group_{t_i}^{vm_j}, S\_VM_{t_i})$ is the number of periods necessary to execute all tasks of $Rec\_Group_{t_i}^{vm_j}$ in the VMs of  $S\_VM_{t_i}$ plus the number of periods necessary to  boot the VMs of $S\_VM_{t_i}$.

The proposed backup scheduling algorithm is presented in Algorithm \ref{alg:backup}, where Table \ref{tab:var2} shows the used variables  and Table \ref{tab:proc2} describes the used  procedures and functions.
As can be seen in line \ref{algB:recg},  $Rec\_Group_{t_i}^{vm_j}$ is created for each task $t_i \in Queue^{vm_j}$. 

This algorithm employs a scheduling strategy similar to that presented in Algorithm \ref{alg:static}, in which tasks are scheduled on different VMs using  a best-fit heuristic. However, unlike the Algorithm \ref{alg:static}, in Algorithm \ref{alg:backup}, the VMs selection prioritizes the on-demand VM with the cheapest monetary cost, resulting from the product of its price and the execution time of a backup task on it.

Note that the backup scheduling has to ensure that if a migration event occurs, the number of periods required to perform the backup tasks respects the deadline. Thus,  the VMs chosen in the function $get\_best\_VM$ (lines \ref{get1} and \ref{get2}) guarantees that $end_{t_i} + runtime(Rec\_Group^{vm_j}_{t_i},  S\_VM_{t_i}) <D $, where $end_{t_i}$ is the end time of the primary task $t_i$ .

After scheduling all backup tasks of $Rec\_Group^{vm_j}_{t_i}$,  the period  when the migration of tasks will have to start to meet the deadline,  $start\_bkp^{vm_j}_{t_i}$, is computed (see lines \ref{compRuntime} and \ref{compStart}).

{
\begin{algorithm}[]
		\caption{\textit{Backup Task Scheduling}}
		\label{alg:backup}
		\small
        \begin{algorithmic}[1]
        \renewcommand{\algorithmicrequire}{\textbf{Input:}}
		\renewcommand{\algorithmicensure}{\textbf{Output:}}
        \REQUIRE  $A\_VM$, $M$, $D$
        \STATE $B\_VM = \emptyset$;

		\FOR{$ \textbf{ all } vm_j \in A\_VM \textbf{ such that } vm_j^{market} = spot $}
		
		    \FOR{$ \textbf{ all } t_i \in Queue^{vm_j}$}
		    
		        \STATE*\textit{/*Create the $Rec\_Group_{t_i}^{vm_j}$ using Equation \ref{eq:error}*/}
		        
		        \STATE $Rec\_Group_{t_i}^{vm_j} = 
		        create\_recovery\_Group(t_i)$ \label{algB:recg}
		        
		        \STATE* \textit{/* Schedule each $t_k \in Rec\_Group_{t_i}^{vm_j}$ on a set of VMs. The VMs choice is guided by the monetary cost resulting from the product of price and execution time */};
		        
		        \STATE $S\_VM_{t_i} = \emptyset$;
		       
		       \FOR{$ \textbf{ all } t_k \in Rec\_Group_{t_i}^{vm_j} $}\label{loopScheduler}
		       
    		       \STATE $vm_{bkp} = NONE$;
    		       \STATE* \textit{/* Select a VM able to execute $t_k$, without violating the  deadline, with the smallest monetary cost*/}
    		       \STATE $vm_{bkp} = get\_best\_VM(t_k, S\_VM_{t_i})$;\label{get1}
    		       
    		       \IF{$ vm_{bkp} \textbf{ is } NONE $}
        		        \STATE $vm_{bkp} = get\_best\_VM(t_k, M)$;\label{get2}
    		       \ENDIF
    		       
    		       \STATE $insert(t_k, vm_{bkp})$;
    		       
    		       \STATE $update(S\_VM_{t_i}, vm_{bkp})$;
		       \ENDFOR
		  
		    \STATE $rtime  = runtime(Rec\_Group_{t_i}^{vm_j}, S\_VM_{t_i})$;\label{compRuntime}
		    \STATE $start\_bkp_{t_i}^{vm_j} = compute\_stbkp(rtime, D)$; \label{compStart}
            \STATE $update(B\_VM, S\_VM_{t_i})$;
        \ENDFOR
    \ENDFOR
    \STATE $ BQ = create\_backup\_map(B\_VM)$;
    \end{algorithmic}
\end{algorithm}
 }

 \begin{table}[htbp]
\caption{Variables of  Backup Scheduling Heuristic}
\begin{center}
\begin{tabular*}{\linewidth}{|c|l|l}
\hline
\textbf{Name} & \multicolumn{2}{|l|}{\textbf{\textit{Description}}} \\
\hline
\begin{tabular}[c]{@{}l@{}}$M$ \end{tabular} & \multicolumn{2}{p{6cm}|}{Set of  VMs} \\
\hline
\begin{tabular}[c]{@{}l@{}}$D$\end{tabular} & \multicolumn{2}{p{6cm}|}{Deadline defined by the user, to be respected even in presence of VM hibernation}\\
\hline
\begin{tabular}[c]{@{}l@{}}$A\_VM$\end{tabular}& \multicolumn{2}{p{6cm}|}{Set of VMs selected to execute primary tasks} \\
\hline
\begin{tabular}[c]{@{}l@{}}$B\_VM$ \end{tabular}& \multicolumn{2}{p{6cm}|}{Set of VMs selected to execute backup tasks} \\
\hline
\begin{tabular}[c]{@{}l@{}}$vm_j$ \end{tabular} & \multicolumn{2}{p{6cm}|}{VM selected to execute primary tasks} \\
\hline
\begin{tabular}[c]{@{}l@{}}$Queue^{vm_j}$\end{tabular} & \multicolumn{2}{p{6cm}|}{Set of  tasks scheduled on $vm_j$}\\
\hline
\begin{tabular}[c]{@{}l@{}}$Rec\_Group_{t_i}^{vm_j} $\end{tabular} & \multicolumn{2}{p{6cm}|}{Set of backup tasks  to be executed due to hibernation of $vm_j$ along $t_i$ execution}\\
\hline
\begin{tabular}[c]{@{}l@{}}$S\_VM_{t_i}$\end{tabular} & \multicolumn{2}{p{6cm}|}{Set of VMs selected to execute backup tasks of the $Rec\_Group^{vm_j}_{t_i}$}  \\
\hline
\begin{tabular}[c]{@{}l@{}}$t_k$\end{tabular} & \multicolumn{2}{p{6cm}|}{A backup task}  \\
\hline
\begin{tabular}[c]{@{}l@{}}$vm_{bkp}$\end{tabular} & \multicolumn{2}{p{6cm}|}{VM selected to execute backup task $t_k$}  \\
\hline
\begin{tabular}[c]{@{}l@{}}$rtime$\end{tabular} & \multicolumn{2}{p{6cm}|}{Number of periods required to execute all tasks of  $Rec\_Group_{t_i}^{vm_j}$ using VMs of $S\_VM_{t_i}$}  \\
\hline
\begin{tabular}[c]{@{}l@{}}$start\_bkp^{vm_j}_{t_i}$\end{tabular} &  \multicolumn{2}{p{6cm}|}{Time when the migration of tasks in   $vm_j$  must start due to hibernation along $t_i$ execution} \\
\hline
\begin{tabular}[c]{@{}l@{}}$BQ$\end{tabular} &  \multicolumn{2}{p{6cm}|}{Scheduling map of backup tasks  containing  VMs  and the  starting times of  backup tasks} \\
\hline

\end{tabular*}
\label{tab:var2}
\end{center}
\end{table}

 \begin{table}[htb]
\caption{Functions and Procedures of Backup Scheduling Heuristic}
\begin{center}
\begin{tabular*}{\linewidth}{|c|l|l}
\hline
\textbf{Name} & \multicolumn{2}{|l|}{\textbf{\textit{Description}}} \\
\hline
\begin{tabular}[c]{@{}l@{}}\\$create\_recovery\_Group(t_i)$\end{tabular} & \multicolumn{2}{p{3.4cm}|}{Create $Rec\_Group_{t_i}^{vm_j}$ using Equation \ref{eq:error}} \\
\hline
\begin{tabular}[c]{@{}l@{}}\\$get\_best\_VM(t_k, S\_VM_{t_i})$\end{tabular} & \multicolumn{2}{p{3.4cm}|}{Select a VM of $S\_VM_{t_i}$  that executes  $t_k$ with minimum monetary cost} \\
\hline
\begin{tabular}[c]{@{}l@{}}$ insert(t_k, vm_{bkp})$ \end{tabular}& \multicolumn{2}{p{3.04cm}|}{Insert backup task $t_k$ into the execution queue of $vm_{bkp}$} \\
\hline
\begin{tabular}[c]{@{}l@{}}\\$runtime(Rec\_Group_{t_i}^{vm_j}, S\_VM_{t_i})$\end{tabular} & \multicolumn{2}{p{3.4cm}|}{Calculate the number of periods necessary to execute all tasks in $Rec\_Group_{t_i}^{vm_j}$ using VMs of  $S\_VM_{t_i}$ } \\
\hline
\begin{tabular}[c]{@{}l@{}}\\$compute\_stbkp(rtime, D) $ \end{tabular}& \multicolumn{2}{p{3.4cm}|}{Calculate the time when the migration procedure  has to start to execute  backup tasks for  $rtime$  periods respecting $D$ } \\
\hline
\begin{tabular}[c]{@{}l@{}}\\$ update(B\_VM, S\_VM_{t_i})$ \end{tabular}& \multicolumn{2}{p{3.4cm}|}{Include VMs of $S\_VM\_{t_i}$ into  $B\_VM$} \\
\hline
\begin{tabular}[c]{@{}l@{}}\\$ create\_backup\_map(B\_VM)$ \end{tabular}& \multicolumn{2}{p{3.4cm}|}{Create the scheduling map of backup tasks} \\
\hline
\end{tabular*}
\label{tab:proc2}
\end{center}
\end{table}

\section{Experimental Results}
\label{sec:results}

This section presents execution times and monetary costs  of  simulations accomplished  with  real BoT applications,  using the configuration of Amazon EC2 virtual machines, and  considering a real VMs market history.

According to the information on  Amazon Web Server (AWS)\footnote{\url{https://docs.aws.amazon.com/AWSEC2/latest/UserGuide/spot-interruptions.html}}, only the VMs of families C3, C4, C5, M4, M5, R3, and R4 with memory below  100 GB, running in the spot market, are able to hibernate if an interrupt occurs. Therefore, for the purposes of this work, the fourth generation general purpose VMs (M4) and the third and fourth generation VMs optimized for computation (C3 and C4) were used. By choosing the third and fourth generation VMs, it was possible to compute the  slowdown using the data from
\cite{gillam2014should}.

The workload used in the  evaluation were obtained from \cite{reiss2011google}, a database that contains the execution traces of jobs  submitted to Google's servers throughout the month of March 2011. Based on these traces, we have defined: (i) the number of tasks of a job; (ii) the execution time of each task of the job; and (iii) the average memory footprint. For the experiments, four BoT-type jobs  were chosen from the first 10 days of the traces. Table \ref{BotCharacteristics} summarizes the main characteristics of these jobs, followed by the corresponding  deadlines considering the virtual machines used in our tests.
We adopted the shortest deadlines which enable the generation of valid primary and backup scheduling, for each job. These values were computed iteratively, using $VM\_time_{limit}$  value equals to 0.5, starting with $D = 1(h)$ in increments of 1 hour, stopping at the first valid scheduling given by the algorithms \ref{alg:static} and \ref{alg:backup}.
The execution times were  obtained  from Google machines  used in  2011. As the hardware information and computational capacity of these machines are not provided, we assumed that these times were obtained with the VM with the lowest computational power, whose memory capacity was sufficient to meet the requirements of the tasks. As we can observe in  Table \ref{VMSCharacteristics}, among the VMs, the ones containing  VCPUs with the lowest computational power are c3.large and m4.large. Therefore, they are considered our baseline regarding processing capacity.
Spot and on-demand VM prices were obtained on September 10, 2018, considering us-east-1 and us-east-1a regions.
Table \ref{VMSCharacteristics}  shows the characteristics of these VMs, along with the corresponding slowdown values of their VCPUs. Based on the latter and considering, as mentioned above, that the duration of the tasks, extracted from Google traces, were obtained from execution them on the slowest VMs (base time duration),  the duration of each task in the other VMs was obtained through the product of the respective slowdown value by its base duration.

\begin{table}[!htbp]
\centering%
\caption{BoT attributes and deadlines\label{BotCharacteristics}}
\begin{tabularx}{\linewidth}{@{}|c| *8{>{\centering\arraybackslash}X|}@{}}
\hline
\multirow{2}{*}{\textbf{Job ID}} & 
\multirow{2}{*}{\textbf{$\#$Tasks}} &
\multirow{2}{*}{\textbf{Memory}}&
\multicolumn{3}{|c|}{\textbf{Execution time of a task }} &
\multirow{2}{*}{\textbf{$D$}} \\
\cline{4-6}
&&&min. & avg. & max. &\\
\hline
J207   & 31 & 6.10 GB   & 7.03 (h)         & 19.75 (h)        & 49.31 (h)        & 17.00 (h)     \\ \hline
J402   & 103& 2.90 GB   & 7.87 (h)         & 29.83 (h)        & 94.04 (h)        & 29.00 (h)      \\ \hline
J819   & 68& 3.97 GB    & 6.42 (h)         & 18.87 (h)        & 51.53 (h)        & 16.00  (h)      \\ \hline
J595   & 97& 3.14 GB    & 7.15 (h)         & 45.80 (h)        & 120.39 (h)       & 40.00 (h)\\ 
\hline
\end{tabularx}
\end{table}

\begin{table}[!htbp]
\centering%
\caption{VMs attributes \label{VMSCharacteristics}}
\begin{tabularx}{\linewidth}{@{}|c| *6{>{\centering\arraybackslash}X|}@{}}
\hline
\textbf{Type} & \textbf{\#VCPUs} & \textbf{Memory} & \textbf{On-demand price} & \textbf{Spot price} & \textbf{slowdown}\\
\hline
m4.large		& 2		& 8.00 GB	& 0.1\$	   & 0.0324\$  & 1.000\\
\hline
c3.large		& 2		& 3.75 GB	& 0.10\$   & 0.0294\$  & 1.000\\
\hline
c4.large		& 2		& 3.75 GB	& 0.1\$	   & 0.0308\$  & 0.655\\
\hline
m4.2xlarge		& 8		& 32.0 GB	& 0.4\$	   & 0.1326\$  & 0.672\\
\hline
m4.xlarge		& 4		& 16.0 GB	& 0.2\$	   & 0.0648\$  & 0.477\\
\hline
m4.4xlarge		& 16	& 64.0 GB	& 0.8\$	   & 0.3257\$  & 0.513\\
\hline
c3.xlarge		& 4		& 7.50 GB	& 0.21\$   & 0.0588\$  & 0.323\\
\hline
c4.xlarge		& 4		& 7.50 GB	& 0.199\$  & 0.0617\$  & 0.332\\
\hline
c3.2xlarge		& 8		& 15.0 GB	& 0.42\$   & 0.1175\$  & 0.475\\
\hline
c4.2xlarge		& 8		& 15.0 GB	& 0.398\$  & 0.1262\$  & 0.447\\
\hline
c3.4xlarge		& 16	& 30.0 GB	& 0.84\$   & 0.2350\$  & 0.163\\
\hline
c4.4xlarge		& 16	& 30.0 GB	& 0.796\$  & 0.2535\$  & 0.162\\
\hline
c4.8xlarge		& 36	& 60.0 GB	& 1.591\$  & 0.4986\$  & 0.162\\
\hline
c3.8xlarge		& 32	& 60.0 GB	& 1.68\$   & 0.4700\$  & 0.161\\
\hline
\end{tabularx}
\end{table}

\subsection{Experimental results in different hibernation scenarios}\label{sub:scenarios}
In order to evaluate the effectiveness of our scheduling solution in terms of makespan and monetary cost, 
 we compared it with an strategy ({\it On-demand}) that uses only on-demand virtual machines, while for evaluating the impact of hibernation, we compared it with a strategy that migrates tasks as soon as the VM, where the tasks have been allocated, hibernates ({\it Immediate Migration}), i.e., the latter does not consider the possibility that the VM might resume. Furthermore, we also consider two possible scenarios of execution of our scheduling:  (1) no spot VM hibernates ({\it No Hibernation}) and (2) a spot VM hibernates and, in this case, either the tasks need to be migrated ({\it Hibernation with Migration}) or the VM resumes in time to not violate deadline ({\it Hibernation}).
In case of hibernation, the latter initiates two hours after the job  starts. For the {\it Hibernation with Migration} execution,  the duration of hibernation is set to 1000 hours, thus forcing task migration; for {\it Hibernation}, hibernation duration is just 3 hours and, therefore, task migration is not carried out.
Aiming a more accurate analysis of the results, only one spot VM can hibernate in  {\it Hibernation} and {\it Hibernation with Migration} executions. In addition, only one backup migration takes place in {\it Hibernation with Migration} execution. Finally, the experiments randomly select the spot VM that should hibernate.

\subsubsection{Hibernation without Migration}

Figures \ref{fig:costwithoutMigration} presents the monetary costs in the {\it Hibernation} scenario, i.e., our scheduling does not  migrate tasks because the hibernated spot VM instance resumes in time to meet the application's deadline.
As we can observe in the figure, its monetary cost is similar to the one without hibernation, represented by the {\it Hibernation} and {\it No Hibernation} bars respectively.
Such a result is expected since, according to the new pricing policy defined by AWS in December 2017, the user only pays for the time the  spots are running, and during hibernation, the user is charged only for storage, whose price on September 10, 2018 was  $0.10$ per GB per month. As, in the experiments, the maximum hibernation time is shorter than 30 hours, it is, thus,  negligible. When  our solution is compared  with the one that migrates tasks as soon as the hibernation occurs ({\it Immediate Migration}), we observe that the latter is more expensive than the second in 59.97\%, 26.74\%, 55.15\% e 40.51\%, for J207, J402, J819 and J595, respectively. Such a difference in price can be explained since in the {\it Immediate Migration}, the user was charged for the two hours of execution of the spot VMs as well as for the on-demand VMs used for migration. On the other hand, it is worth mentioning that the {\it Immediate Hibernation} monetary cost is, for the four jobs,  on average, 57.31\% lower than the ({\it On-demand}) one. This happens because part of the tasks were executed as primary ones in spot VM with high computational power, and, therefore, fewer slots were needed to complete execution on the on-demand VMs.

\begin{figure}[!htb]
    \centering
    \includegraphics[scale=0.20]{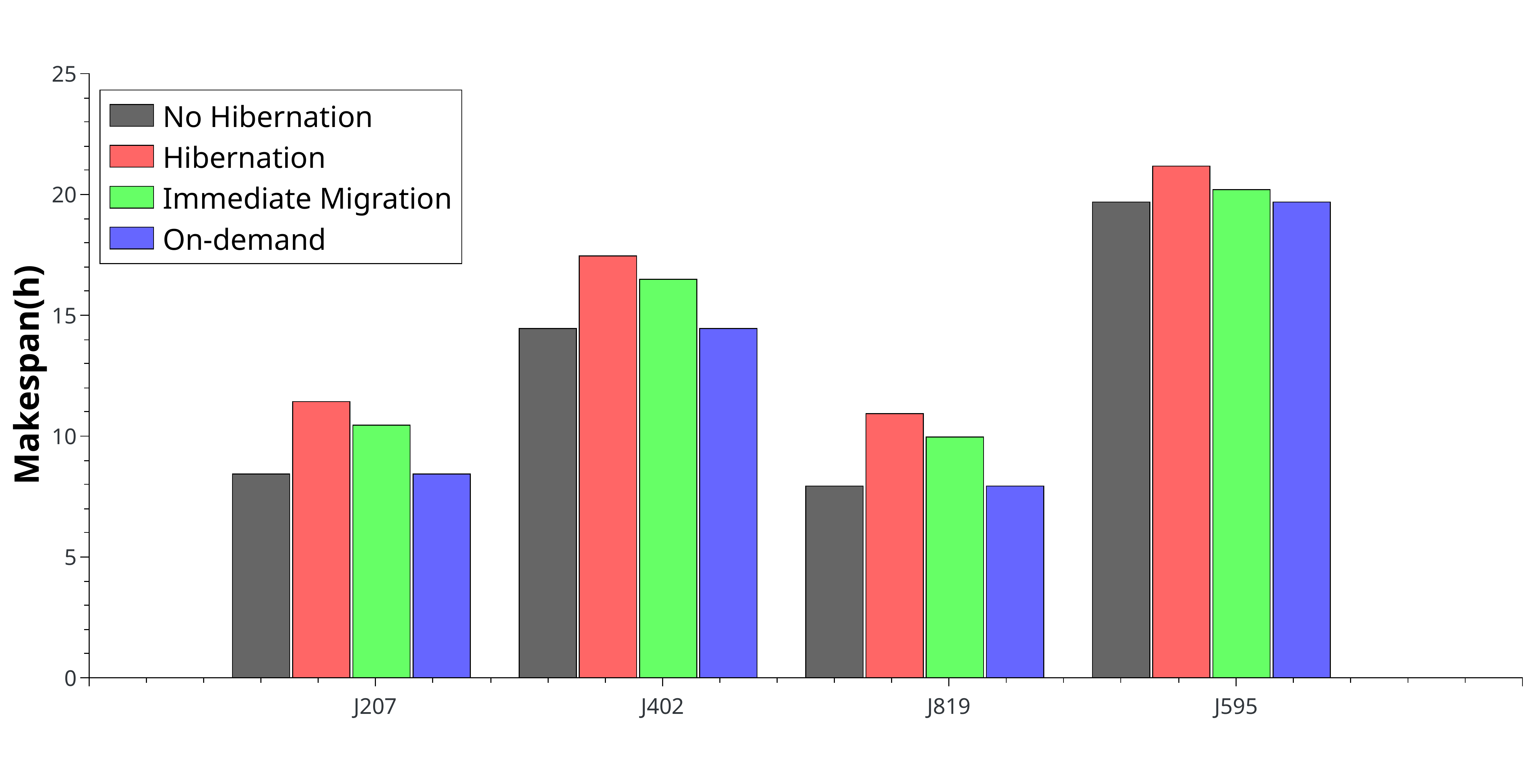}
    \caption{Makespans considering VM the spot VM resumes.}
    \label{fig:makespanwithoutMigration}
\end{figure}

Figure \ref{fig:makespanwithoutMigration} shows  that our solution, {\it Hibernation}, has a makespan longer than the {\it Immediate Migration} strategy. This occurs because the former has an additional 3 hours due to hibernation, while the latter migrates immediately, continuing running the job's tasks within this 3 hours.

In contrast, since the VMs usually chosen by the {\it Immediate Migration} strategy are low cost ones, performing poorly, its makespan can be longer than {\it On-demand} and  {\it No Hibernation} ones, that allocate VMs with higher computation power. Moreover,  when the virtual machine hibernates, the executing task  is re-started from the beginning in another VM.  So, its execution time can be computed (in the makespan) almost twice in the worst case.

\subsubsection{Hibernation with Migration}

\begin{figure}[!htb]
    \centering
    \includegraphics[scale=0.20]{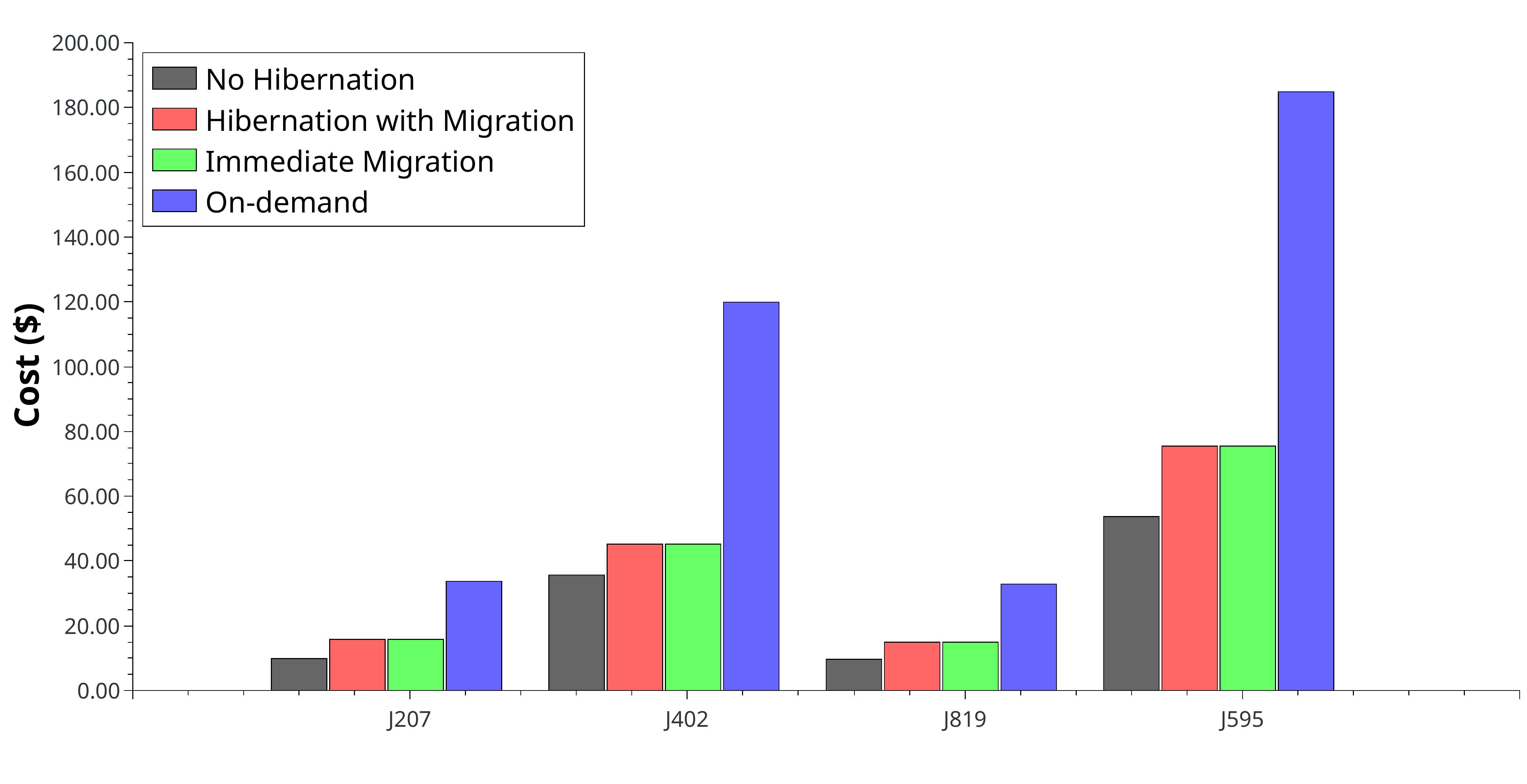}
    \caption{Monetary costs with tasks migration.}
    \label{fig:costWithmigration}
\end{figure}

\begin{figure}[!htb]
    \centering
    \includegraphics[scale=0.20]{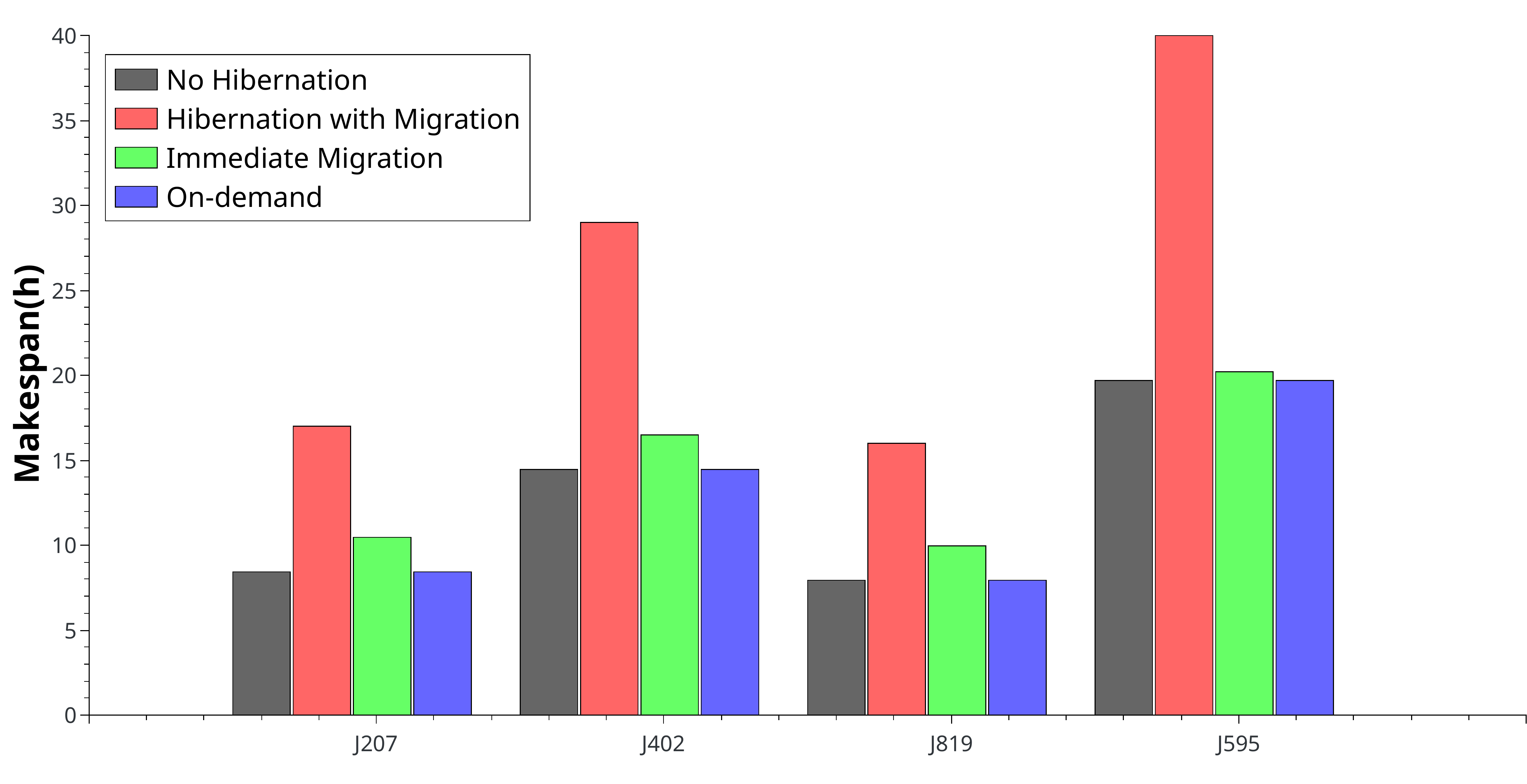}
    \caption{Makespans with tasks migration.}
    \label{fig:makepanwithMigration}
\end{figure}

Table \ref{usedVMs} presents the  number of VMs and the corresponding types used before and after migration for each job, where the hibernated VM is indicated by (H). Note that we consider that only one VM hibernates in these tests, i.e., even if  several instances of a same VM type are allocated, only one of them may hibernate.  Figures \ref{fig:costWithmigration} and \ref{fig:makepanwithMigration}  respectively show the monetary costs  and  makespans in the scenario   where our scheduling ({\it Hibernation with Migration}) migrates tasks of a hibernated spot VM.

\begin{table}[!htbp]
\centering%
\caption{VMs used to execute jobs before and after migration\label{usedVMs}}
\begin{tabularx}{\linewidth}{@{}|c| *2{>{\raggedright\arraybackslash}X|}@{}}
\hline
\textbf{Job ID} & \textbf{Before Migration} & \textbf{After Migration} \\
\hline
J207  & (H) 1-c3.4xlarge(spot) 1-c3.8xlarge(spot)  1-c4.8xlarge(spot) 	& 1-c3.4xlarge(on-demand) 1-c3.8xlarge(spot)  1-c4.8xlarge(spot) \\ \hline
J402  & (H) 5-c3.4xlarge(spot) 3-c4.8xlarge(spot) 						& 1-c3.4xlarge(on-demand) 4-c3.4xlarge(spot) 3-c4.8xlarge(spot)    \\ \hline
J819  & 1-c3.4xlarge(spot) (H)1-c3.8xlarge(spot) 1-c4.8xlarge(spot)	&  1-m4.4xlarge(on-demand) 1-c3.4xlarge(on-demand) 1-c3.4xlarge(spot) 1-c4.8xlarge(spot) \\ \hline
J595  & 2-c4.8xlarge(spot)  (H)3-c3.8xlarge(spot) 2-c3.4xlarge(spot)	&  2-c4.8xlarge(on-demand) 2-c4.8xlarge(spot) 2-c3.8xlarge(spot) 2-c3.4xlarge(spot)   \\ 
\hline
\end{tabularx}
\end{table}

The monetary  cost of {\it Hibernation with Migration} strategy is equal to the {\it Immediate Migration} since the backup map used by both of them are similar. These costs are higher (on average, 30.74\% in our experiments) than the one required by the primary scheduling alone ({\it No Hibernation}), since the former use spot VMs within the first two hours of execution,  as well as  on-demand VMs  for backup migration. On the other hand they are 136.00\% lower than {\it On-demand} strategy costs.
In terms of makespan, the  {\it Hibernation with Migration} makespans is close to the deadlines defined in the Table \ref{BotCharacteristics}. Such a behaviour is expected since our approach waits till the $start\_bkp$, which is the latest time that hibernation can be tolerated without exceeding the deadline. Note that in the case of the {\it Immediate Migration} strategy, the makespan  is shorter than the {\it Hibernation with Migration} one. In our experiments, this difference was, on average, 74.26\%. 

Note that, although in some cases, the tasks  can migrate  to VMs  of equivalent processing powers  (see the case of J207), even in {\it Immediate Migration} strategy, the makespan  increases.  As pointed out in the previous section, it happens because  the  execution time of a task, initially started in  a VM that hibernates along its execution, can be computed almost twice, when it migrates, in the worst case.

\begin{figure}[!htb]
    \centering
    \includegraphics[scale=0.20]{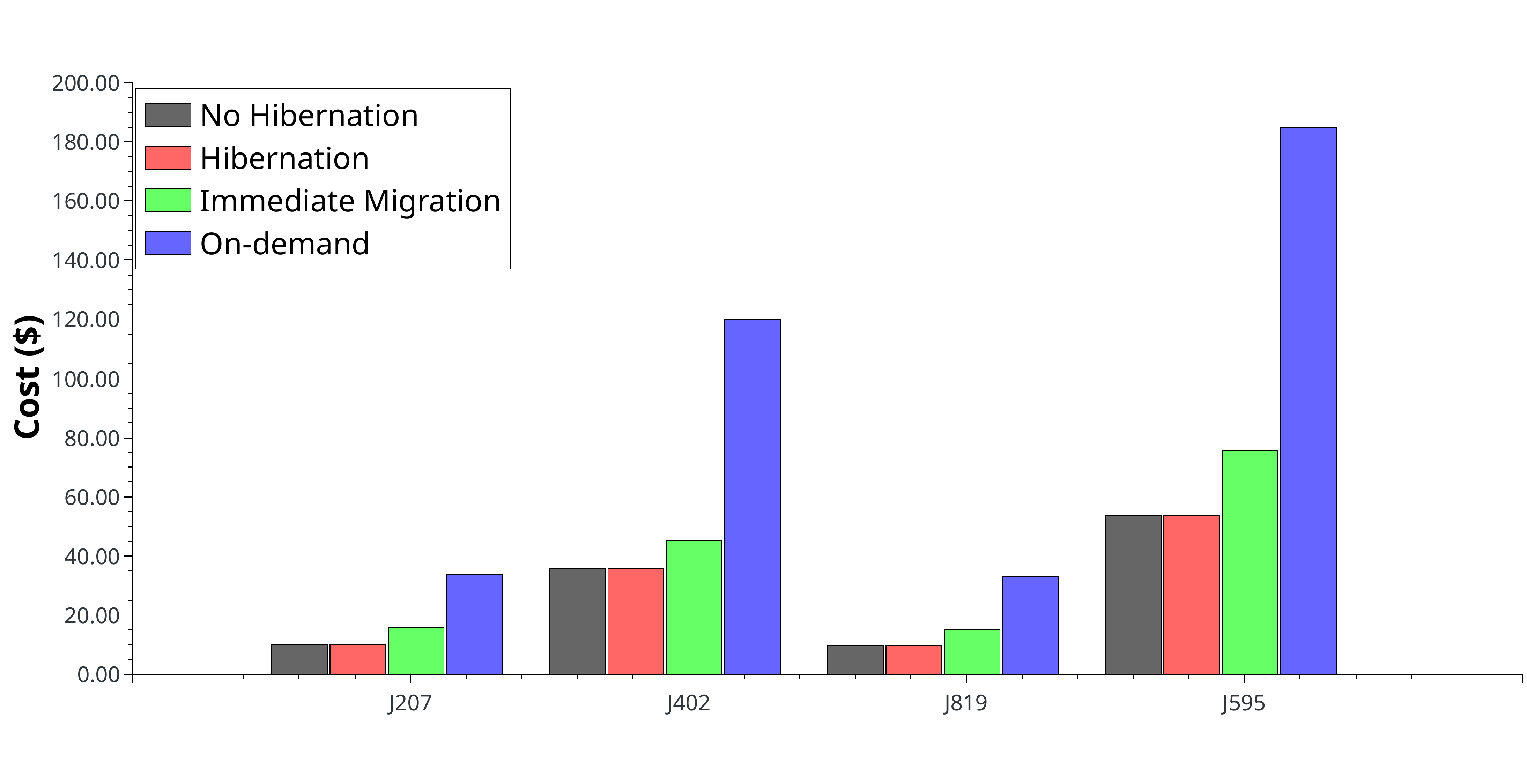}
    \caption{Monetary costs considering that the spot VM resumes.}
    \label{fig:costwithoutMigration}
\end{figure}

When comparing {\it Hibernation} with {\it Hibernation with Migration}, the duration of hibernation has an impact in both makespans due to the duration of the execution itself. However, in the case of {\it Hibernation}, where the hibernated spot VM resumes in time to respect the deadline, the monetary cost is lower than {\it Hibernation with Migration}, as we can confirm in Figures \ref{fig:costwithoutMigration} and \ref{fig:costWithmigration}.

\subsection{Experimental Results  with hibernation based on the variation of spots price}

\begin{figure}[!htb]
    \centering
    \includegraphics[scale=0.18]{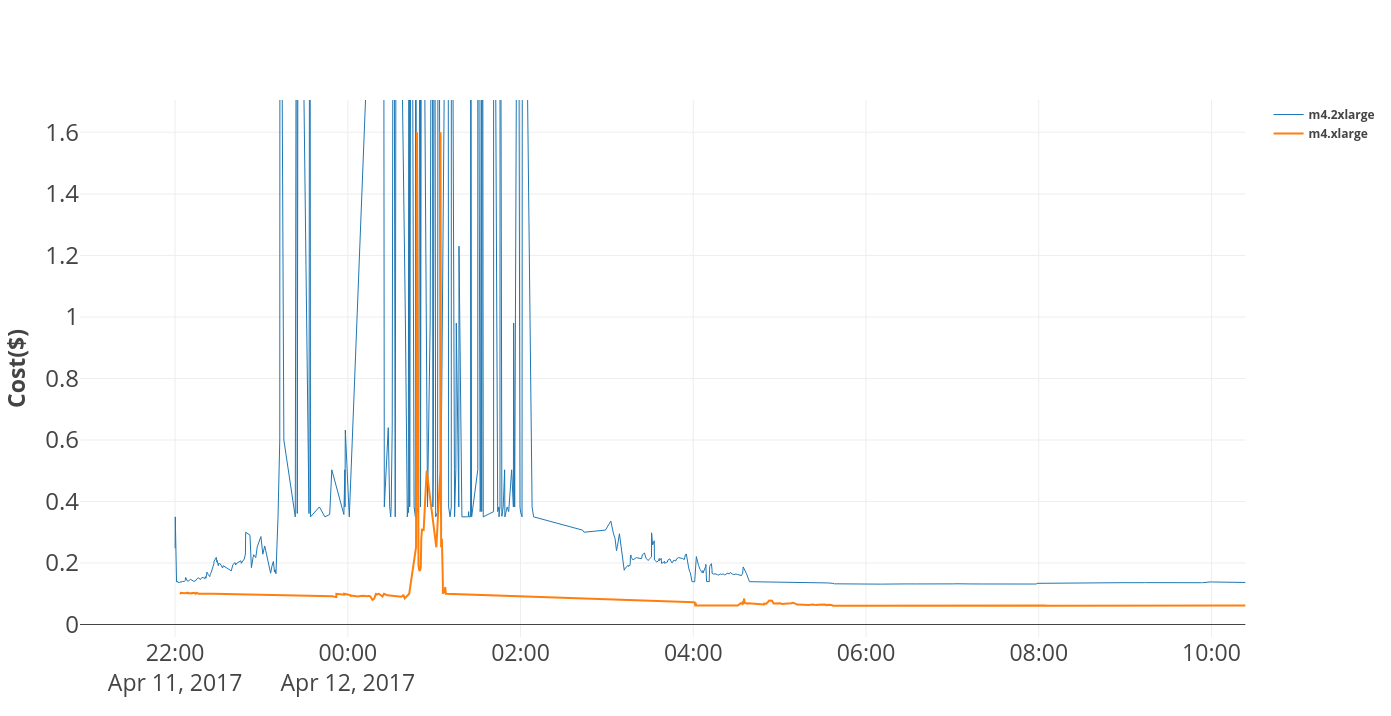}
    \caption{Sample of the price variation of two EC2 VMs in the spot market during april 2017.}
    \label{fig:priceVariation}
\end{figure}

 The results presented in this section are from experiments that consider spot  price variations  for regions us-east-1 and zone us-east-1a between March and April of 2017, defining hibernation traces for the VMs of Table \ref{VMSCharacteristics}. 
 That history of price variation predates the changes in AWS  pricing policies, occurred in December 2017, which stabilized the prices of VMs such that peaks of variation ceased to occur\footnote{\url{https://aws.amazon.com/blogs/compute/new-amazon-ec2-spot-pricing/}}. As shown in Figure \ref{fig:priceVariation}, in the previous policy, prices could have significant peaks of variation, with intervals lasting a few minutes or hours.

The hibernation traces were generated considering a fixed threshold of 
$\$ 0.4$, which represents the average price value in the first 24 hours of the history. Thus, the onset of hibernation is the period in which the VM price is higher than this value. Analogously, when the price drops to a value below the threshold, we consider that the VM resumes execution.
The generated traces has two hibernation points: (1) c4.8xlarge VMs hibernation at $4.21$ hours after the start of execution and lasting $43.51$ minutes; (2) c3.4xlarge VMs hibernation at $23.7$ minutes after the start of execution and lasted $1.22$ hours. 

The number of VMs affected by hibernation is not the same for all evaluated jobs. While in the J207 and J819 jobs only 2 VMs hibernate, in Job J402 there are 8 hibernations of different VMs. This variation is expected, since different job tasks are scheduled to VMs of different types.

\begin{figure}[!htb]
    \centering
    \includegraphics[scale=0.20]{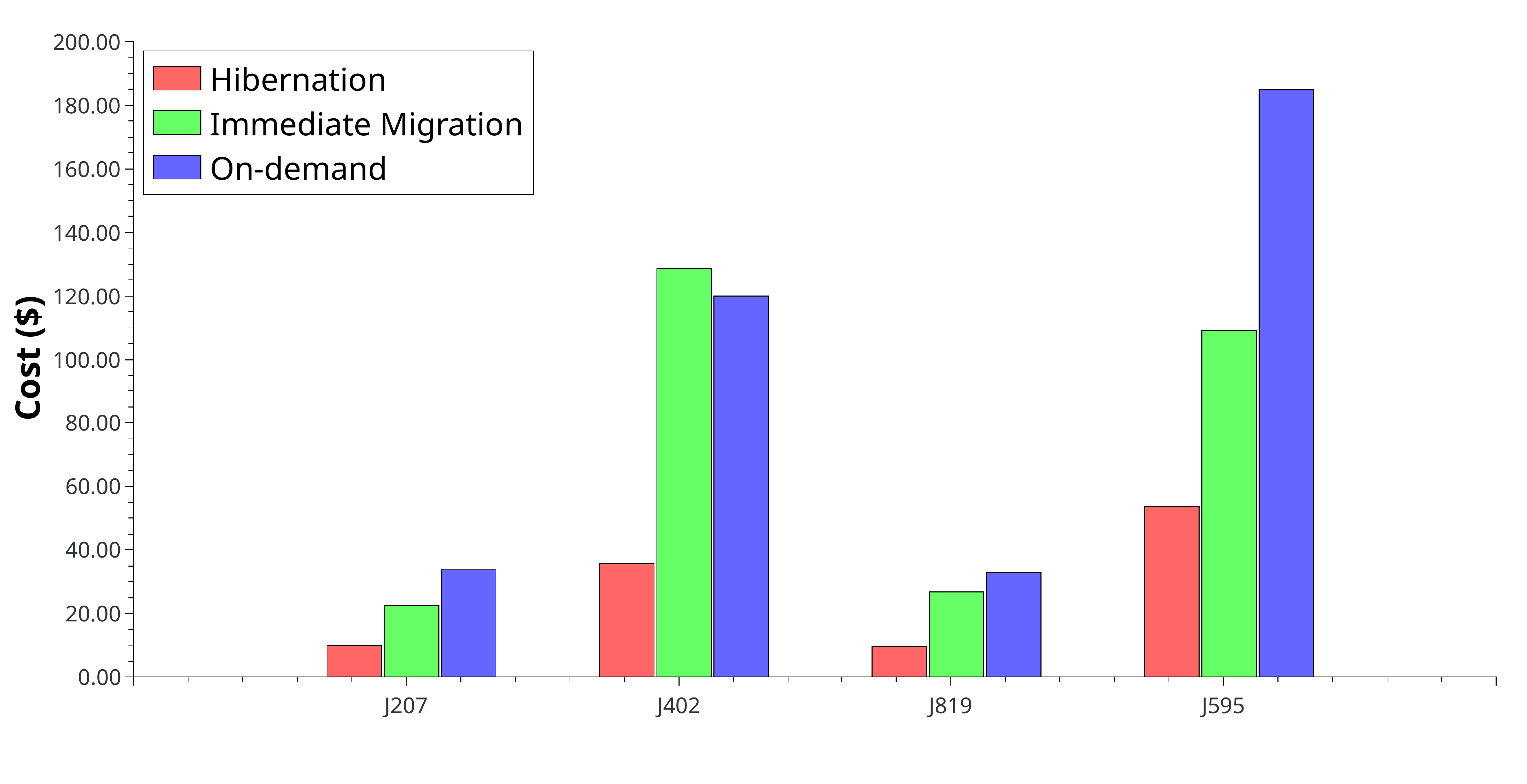}
    \caption{Monetary costs with hibernation based on AWS price history.}
    \label{fig:aws_cost}
\end{figure}

As can be observed in Figure \ref{fig:aws_cost}, our solution, {\it Hibernation}, presents the lowest monetary cost, with an average difference of 167.43 \%, relative to the {\it Immediate Migration}'s one, and 240.94 \% in relation to the {\it  On-demand}'s  one.
It is noteworthy that in Job J402, {\it Immediate Migration} has a cost which is  6.73 \% higher than the {\it  On-demand}'s one. The former used 8 on-demand VMs for migration, which raised the monetary cost,  added to  the costs of the VMs spots used until the beginning of their hibernation. On the other hand, for Job J402, our approach presents a significantly lower cost than  the {\it Immediate Migration}'s one (260.26 \%), since the duration of none of the of hibernation of the corresponding spot VMs triggered the migration of their tasks.

\begin{figure}[!htb]
    \centering
    \includegraphics[scale=0.20]{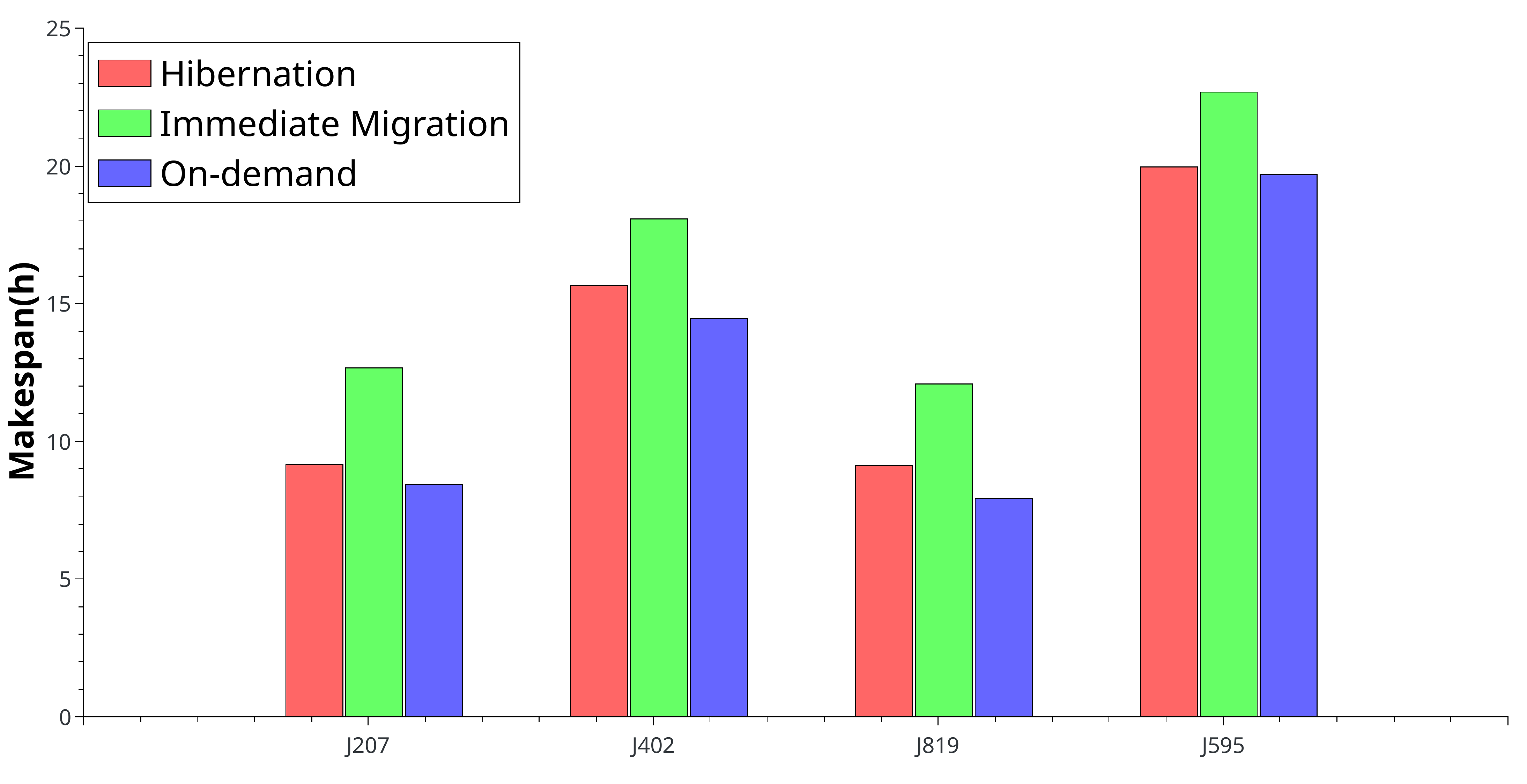}
    \caption{Makespans with hibernation based on AWS price history.} 
    \label{fig:aws_makespan}
\end{figure}

Regarding makespan, shown in Figure \ref{fig:aws_makespan}, our approach is 7.52 \%   longer  than  {\it On-demand}'s one  and 24.92 \% shorter than {\it Immediate Migration}'s one. These difference can be explained since the duration of VMs hibernation is up to  $ 1.22 $ hours, it is not necessary to start the task migration process in any of the evaluated jobs. Therefore, this increase is due only to the hibernation of the VMs. On the other hand, in {\it Immediate Migration} the scheduling of backup tasks chooses firstly cheaper on-demand VMs, usually  with lower computational power.

Although our approach increases the makespan when compared {\it  On-demand}'s one, the monetary costs are lower than the two other approaches. Thus, the results from the experiments with the hibernation trace confirm  those from the previous experiments.

\newpage
\section{Concluding Remarks and Future Work}
~\label{sec:conclusion}

 This paper proposed a static scheduling for bag-of-task applications  with deadline constraints, using both hibernation-prone spot VMs (for cost sake) and on-demand VMs.  Our scheduling  aims at  minimizing  monetary costs of bag-of-tasks, respecting application's deadline and avoiding temporal failures.  Although we have theoretically evaluated the proposed strategy, the characteristics of Bot applications and VMs, as well as the VM market price variation,  were acquired from real scenarios. Our results confirmed the effectiveness of our scheduling and that it tolerates temporal failures.
  
 Short-term directions of our work comprise the automation of the computing of both the minimum deadline and the $VM\_time_{limit}$,  in accordance with the characteristics of the application and  the available  virtual machines as well as the maximum number of hibernations tolerated at each spot virtual machine. Thus, the user will always have a feasible static scheduling for the  expected scenario.
In longer-term future, we also intend to work in a dynamic  version of the proposed scheduling which periodically takes  checkpoints of the tasks, so that, in the migration case, the tasks can start their executions from the last checkpoints, instead of being re-started from the beginning.

\bibliographystyle{IEEEtran}
\bibliography{IEEEabrv,ref}

\begin{thebibliography}{10}
\providecommand{\url}[1]{#1}
\csname url@samestyle\endcsname
\providecommand{\newblock}{\relax}
\providecommand{\bibinfo}[2]{#2}
\providecommand{\BIBentrySTDinterwordspacing}{\spaceskip=0pt\relax}
\providecommand{\BIBentryALTinterwordstretchfactor}{4}
\providecommand{\BIBentryALTinterwordspacing}{\spaceskip=\fontdimen2\font plus
\BIBentryALTinterwordstretchfactor\fontdimen3\font minus
  \fontdimen4\font\relax}
\providecommand{\BIBforeignlanguage}[2]{{%
\expandafter\ifx\csname l@#1\endcsname\relax
\typeout{** WARNING: IEEEtran.bst: No hyphenation pattern has been}%
\typeout{** loaded for the language `#1'. Using the pattern for}%
\typeout{** the default language instead.}%
\else
\language=\csname l@#1\endcsname
\fi
#2}}
\providecommand{\BIBdecl}{\relax}
\BIBdecl

\bibitem{SmallenCB01}
S.~Smallen, H.~Casanova, and F.~Berman, ``Applying scheduling and tuning to
  on-line parallel tomography,'' in \emph{Proceedings of the 2001 {ACM/IEEE}
  conference on Supercomputing, Denver, CO, USA, November 10-16, 2001,
  {CD-ROM}}, 2001, p.~12.

\bibitem{CasanovaOBW00}
H.~Casanova, G.~Obertelli, F.~Berman, and R.~Wolski, ``The apples parameter
  sweep template: User-level middleware for the grid,'' in \emph{Proceedings
  Supercomputing 2000, November 4-10, 2000, Dallas, Texas, {USA.} {IEEE}
  Computer Society, {CD-ROM}}, 2000, p.~60.

\bibitem{MascagniL03}
M.~Mascagni and Y.~Li, ``Computational infrastructure for parallel,
  distributed, and grid-based monte carlo computations,'' in \emph{Large-Scale
  Scientific Computing, 4th International Conference, {LSSC} 2003, Sozopal,
  Bulgaria, June 4-8, 2003, Revised Papers}, 2003, pp. 39--52.

\bibitem{WhiteTW90}
S.~W. White, D.~C. Torney, and C.~C. Whittaker, ``A parallel computational
  approach using a cluster of {IBM} {ES/3090} 600js for physical mapping of
  chromosomes,'' in \emph{Proceedings Supercomputing '90, New York, NY, USA,
  November 12-16, 1990}, 1990, pp. 112--121.

\bibitem{GSW2015}
\BIBentryALTinterwordspacing
A.~Goder, A.~Spiridonov, and Y.~Wang, ``Bistro: Scheduling data-parallel jobs
  against live production systems,'' in \emph{2015 {USENIX} Annual Technical
  Conference ({USENIX} {ATC} 15)}.\hskip 1em plus 0.5em minus 0.4em\relax Santa
  Clara, CA: {USENIX} Association, 2015, pp. 459--471. [Online]. Available:
  \url{https://www.usenix.org/conference/atc15/technical-session/presentation/goder}
\BIBentrySTDinterwordspacing

\bibitem{ThaiVB18}
L.~Thai, B.~Varghese, and A.~Barker, ``A survey and taxonomy of resource
  optimisation for executing bag-of-task applications on public clouds,''
  \emph{Future Generation Comp. Syst.}, vol.~82, pp. 1--11, 2018.

\bibitem{YaoZlHLL14}
M.~Yao, P.~Zhang, Y.~Li, J.~Hu, C.~Li, and X.~Li, ``Cutting your cloud
  computing cost for deadline-constrained batch jobs,'' in \emph{2014 {IEEE}
  International Conference on Web Services, ICWS, 2014, Anchorage, AK, USA,
  June 27 - July 2, 2014}, 2014, pp. 337--344.

\bibitem{ThaiVB14}
L.~Thai, B.~Varghese, and A.~Barker, ``Executing bag of distributed tasks on
  the cloud: Investigating the trade-offs between performance and cost,'' in
  \emph{{IEEE} 6th International Conference on Cloud Computing Technology and
  Science, CloudCom 2014, Singapore, December 15-18, 2014}, 2014, pp. 400--407.

\bibitem{ThaiVB15}
L.~Thai, B.~Varghese, and Barker, ``Task scheduling on the cloud with hard
  constraints,'' in \emph{2015 {IEEE} World Congress on Services, {SERVICES}
  2015, New York City, NY, USA, June 27 - July 2, 2015}, 2015, pp. 95--102.

\bibitem{GoiriJGT10}
I.~Goiri, F.~Juli{\`{a}}, J.~Guitart, and J.~Torres, ``Checkpoint-based
  fault-tolerant infrastructure for virtualized service providers,'' in
  \emph{{IEEE/IFIP} Network Operations and Management Symposium, {NOMS} 2010,
  19-23 April 2010, Osaka, Japan}, 2010, pp. 455--462.

\bibitem{AupyBMRR13}
G.~Aupy, A.~Benoit, R.~G. Melhem, P.~Renaud{-}Goud, and Y.~Robert,
  ``Energy-aware checkpointing of divisible tasks with soft or hard
  deadlines,'' in \emph{International Green Computing Conference, {IGCC} 2013,
  Arlington, VA, USA, June 27-29, 2013, Proceedings}, 2013, pp. 1--8.

\bibitem{PlankensteinerPFKK08}
K.~Plankensteiner, R.~Prodan, T.~Fahringer, A.~Kert{\'{e}}sz, and P.~Kacsuk,
  ``Fault detection, prevention and recovery in current grid workflow
  systems,'' in \emph{Grid and Services Evolution, Proceedings of the 3rd
  CoreGRID Workshop on Grid Middleware, June 5-6, 2008, Barcelona, Spain},
  2008, pp. 1--13.

\bibitem{ZhengVT09}
Q.~Zheng, B.~Veeravalli, and C.~Tham, ``On the design of fault-tolerant
  scheduling strategies using primary-backup approach for computational grids
  with low replication costs,'' \emph{{IEEE} Trans. Computers}, vol.~58, no.~3,
  pp. 380--393, 2009.

\bibitem{WangBZYX15}
J.~Wang, W.~Bao, X.~Zhu, L.~T. Yang, and Y.~Xiang, ``{FESTAL:} fault-tolerant
  elastic scheduling algorithm for real-time tasks in virtualized clouds,''
  \emph{{IEEE} Trans. Computers}, vol.~64, no.~9, pp. 2545--2558, 2015.

\bibitem{Al-OmariSM04}
R.~Al{-}Omari, A.~K. Somani, and G.~Manimaran, ``Efficient overloading
  techniques for primary-backup scheduling in real-time systems,'' \emph{J.
  Parallel Distrib. Comput.}, vol.~64, no.~5, pp. 629--648, 2004.

\bibitem{CirneBSGV07}
W.~Cirne, F.~V. Brasileiro, D.~P. da~Silva, L.~F.~W. G{\'{o}}es, and
  W.~Voorsluys, ``On the efficacy, efficiency and emergent behavior of task
  replication in large distributed systems,'' \emph{Parallel Computing},
  vol.~33, no.~3, pp. 213--234, 2007.

\bibitem{BenoitHR08}
A.~Benoit, M.~Hakem, and Y.~Robert, ``Fault tolerant scheduling of precedence
  task graphs on heterogeneous platforms,'' in \emph{22nd {IEEE} International
  Symposium on Parallel and Distributed Processing, {IPDPS} 2008, Miami,
  Florida USA, April 14-18, 2008}, 2008, pp. 1--8.

\bibitem{LuLWKPHLL13}
S.~Lu, X.~Li, L.~Wang, H.~Kasim, H.~N. Palit, T.~Hung, E.~F.~T. Legara, and
  G.~K.~K. Lee, ``A dynamic hybrid resource provisioning approach for running
  large-scale computational applications on cloud spot and on-demand
  instances,'' in \emph{19th {IEEE} International Conference on Parallel and
  Distributed Systems, {ICPADS} 2013, Seoul, Korea, December 15-18, 2013},
  2013, pp. 657--662.

\bibitem{MenacheSJ14}
I.~Menache, O.~Shamir, and N.~Jain, ``On-demand, spot, or both: Dynamic
  resource allocation for executing batch jobs in the cloud,'' in \emph{11th
  International Conference on Autonomic Computing, {ICAC} '14, Philadelphia,
  PA, USA, June 18-20, 2014.}, 2014, pp. 177--187.

\bibitem{SharmaLGIS15}
P.~Sharma, S.~Lee, T.~Guo, D.~E. Irwin, and P.~J. Shenoy, ``Spotcheck:
  designing a derivative iaas cloud on the spot market,'' in \emph{Proceedings
  of the Tenth European Conference on Computer Systems, EuroSys 2015, Bordeaux,
  France, April 21-24, 2015}, 2015, pp. 16:1--16:15.

\bibitem{SubramanyaGSIS15}
S.~Subramanya, T.~Guo, P.~Sharma, D.~E. Irwin, and P.~J. Shenoy, ``Spoton: a
  batch computing service for the spot market,'' in \emph{Proceedings of the
  Sixth {ACM} Symposium on Cloud Computing, SoCC 2015, Kohala Coast, Hawaii,
  USA, August 27-29, 2015}, 2015, pp. 329--341.

\bibitem{gillam2014should}
L.~Gillam and J.~O'Loughlin, ``Should infrastructure clouds be priced entirely
  on performance? an ec2 case study,'' \emph{International Journal of Big Data
  Intelligence}, vol.~1, no.~4, pp. 215--229, 2014.

\bibitem{reiss2011google}
C.~Reiss, J.~Wilkes, and J.~L. Hellerstein, ``Google cluster-usage traces:
  format+ schema,'' \emph{Google Inc., White Paper}, pp. 1--14, 2011.

\end{thebibliography}

\end{document}